\documentclass{article}
\usepackage{arxiv}

\usepackage{url}
\usepackage{natbib}
\usepackage{lineno}
\usepackage{amsmath}
\usepackage[utf8]{inputenc}
\usepackage{physics}
\usepackage{hyperref}
\usepackage{graphicx}
\usepackage{xcolor}
\usepackage{soul}
\usepackage{comment}
\usepackage{rotating}

\title{SW1D: A Python package for seismic surface wave solutions in 1-D elastic media}

 \author{
  Piyush Kolhe \\
   Indian Institute of Science Education and Research Pune\\
   Pune 411008, India \\
    \And
    Arjun Datta \\
   Department of Earth and Climate Science\\
   Indian Institute of Science Education and Research Pune\\
   Pune 411008, India \\
   \texttt{arjundatta@iiserpune.ac.in} \\
   \And
  Shijil Umer \\
   Department of Meteorology and Geophysics \\
   University of Vienna \\
   Vienna, Austria\\
   \And
  Bivas Das \\
   Observatoire de la C\^ote d'Azur \\
   Universit\'e C\^ote d'Azur \\
   CNRS, IRD, G\'eoazur \\
   Sophia Antipolis, France\\
 }

\begin{document}
\maketitle

\begin{abstract}
	We present SW1D, a Python version of a legacy code for the forward computation of Rayleigh and Love wave dispersion curves as well as synthetic seismograms, in elastic, isotropic, laterally homogeneous media. This code solves the surface wave eigenvalue problem for 1-D media using a propagator-matrix technique and subsequently obtains surface wave excitations, in the frequency domain, for a point source described by its moment tensor. SW1D is validated through rigorous theoretical analyses and benchmarked against a well-established seismological software package. Ultimately, we demonstrate its place within an ecosystem of Python-based seismological tools designed for niche applications such as modelling surface wave propagation across lateral discontinuities, or forward modelling of ambient noise cross-correlations with arbitrary spatial distributions of ambient seismic sources.
\end{abstract}








\section{Introduction}

The theory of surface-wave propagation in isotropic, laterally homogeneous (1-D) Earth models is well understood and several techniques exist for the calculation of surface wave Green's functions. All of them first solve the stress-displacement eigenvalue problem and use the eigenfunctions computed therein to obtain surface-wave radiation for a specified seismic moment tensor. Methods to solve the surface wave eigenvalue problem historically evolved along two dominant lines --- first, the propagator matrix technique which began with the work of \citet{Thomson1950} and \citet{Haskell1953} and the propagator theory of \citet{Gilbert1966}; second, the direct integration or shooting technique which is due to \citet{Takeuchi1972}. The former class of methods approximates a vertically varying medium by a stack of homogeneous layers whilst the latter works with continuous variation of properties with depth. There is also a third class of methods based on direct approximation of eigenfunctions as a finite sum of polynomials or other special functions \citep[e.g.][]{Lysmer1970}, which can be applied to the most general depth dependence of properties i.e.\ continuous variation as well as material interfaces \citep[e.g.][]{Denolle2012}. 

The propagator matrix methods underwent a series of modifications and improvements during the second half of the 20th century \citep[see][]{Buchen1996} to overcome high frequency stability and accuracy problems in the original Thomson-Haskell method. These include the delta matrix method \citep{Pestel1963, Dunkin1965, Thrower1965, Watson1970} and the Schwab-Knopoff approach \citep{Knopoff1964matrix, Schwab1971} among others \citep{AboZena1979, Chapman2003, Liu2010}.

This study presents the Python code package SW1D, which is based on the method of \citet{Gomberg1988}. This two-part code uses a propagator-matrix technique to solve the surface wave eigenvalue problem and, subsequently, computes displacement spectra for point sources specified by their moment tensor. In this report we outline the theory behind SW1D (Section \ref{sec:theory}), demonstrate that it provides accurate solutions (Section \ref{sec:testing}), and highlight some of its unique seismological applications (Section \ref{sec:app}).

\section{Theory and Methods}
\label{sec:theory}

\subsection{Solution of the eigenvalue problem}

SW1D uses a propagator matrix technique to solve the surface wave eigenvalue problem (chapter 7 of \cite{Aki2002} and Appendix \ref{app:theory_details}), retaining the original numerical algorithm of \cite{Gomberg1988}. Key aspects of this workflow include the minor vector propagation method of \citet{Woodhouse1980} and the mode counting algorithm of \cite{Woodhouse1981}, which facilitate fast computations. Since \cite{Gomberg1988} provided all the numerical details, we do not repeat them here. We simply note that once the Love and Rayleigh wave eigenvalue problems are solved to obtain the corresponding phase velocities and eigenfunctions, their group velocities ($U^L$ and $U^R$, respectively) are calculated using the following expressions \citep{Aki2002}:

\begin{align}
    U^L = \frac{I_2^L}{cI_1^L} \\
    U^R = \frac{I_2^R + \frac{I_3^R}{2k}}{cI_1^R}
\end{align}

In the above expressions, $c$ is phase velocity, $k$ is wavenumber and the terms $I^{L/R}_j$ denote energy integrals for Love ($L$) or Rayleigh ($R$) waves. The energy integrals follow standard definitions \citep{Aki2002} and are reproduced in Appendix \ref{app:theory_details} for reference. Throughout this paper we use $L/R$ superscripts to denote quantities corresponding to Love or Rayleigh waves.

\subsection{Incorporation of anelastic attenuation}

Anelasticity is included through two effects. First, a finite quality factor $Q$ makes the medium dispersive, i.e.\ the body-wave velocities become weakly frequency-dependent through the relations \citep{Aki2002}:

\begin{align}
    v_P(\omega) &= v_P(\omega_{\text{ref}}) \left( 1 + \frac{1}{\pi Q_\alpha} \ln{\frac{\omega}{\omega_{\text{ref}}}} \right) \\
    v_S(\omega) &= v_S(\omega_{\text{ref}}) \left(1 + \frac{1}{\pi Q_\beta} \ln{\frac{\omega}{\omega_{\text{ref}}}} \right)
\end{align}

These relations are applied layer by layer. $Q_\alpha$ and $Q_\beta$ are the compressional and shear quality factors and $\omega_{\text{ref}} = 2\pi/t_{\text{ref}}$ is the reference angular frequency set by the model reference period $t_{\text{ref}}$. The Lam\'e parameters used in the eigenvalue problem (in particular, $\mu = \rho v_S^2$) are recomputed from these dispersed velocities at each frequency, so that the phase velocities and eigenfunctions carry the physical dispersion.

Attenuation also causes each surface wave mode to lose amplitude as it propagates. The contribution of a mode to the spectrum at epicentral distance $r$ is multiplied by a factor of $\exp{(-\gamma \, r)}$, where $\gamma = \omega/2UQ$, $U$ is the group velocity and the quality factor $Q^{-1}$ is the energy-weighted average of the layer $Q^{-1}$ values using the energy integrals defined in Appendix \ref{app:theory_details}.

\subsection{Earth flattening}

SW1D uses a flat geometry and the sphericity of the Earth is accounted for by means of an Earth-flattening transformation \citep{Biswas1970,Biswas1972}, which can be represented as follows:
\begin{align}
\begin{split}
    z &= a \ln \left(\frac{a}{r} \right) \\
    v_f(z) &= \frac{a}{r} v_s(r) \\
    \rho_f(z) &= \left( \frac{r}{a} \right)^p \rho_s(r)
\end{split}
\end{align}

Here, $a$ is Earth's radius, $r$ is the depth for a spherical Earth model, $z$ is the depth for the flattened model, subscript $f$ denotes a flattened quantity and subscript $s$ denotes the quantity in the spherical Earth model. The factor $p$ is:

\begin{equation}
    p = \begin{cases}
        2.275 & \text{for Rayleigh waves}\\
        5 & \text{for Love waves}
    \end{cases}
\end{equation}

Since the model is made of layers of finite thickness (with a half-space underlying the bottom-most layer), an average of the factor $a/r$ for a layer is used:

\begin{equation}
    \left\langle \frac{a}{r} \right\rangle = \frac{1}{\Delta z} \int \frac{a}{r} \, dz
\end{equation}

For the half-space,
\begin{align}
\begin{split}
    &v_f = \frac{a}{R} v_s \\
    &\rho_f = \left( \frac{R}{a} \right)^p \rho_s \text{,}
\end{split}
\end{align}
where $R$ is the depth at the end of the lowest layer.

\subsection{Calculation of surface wave spectra for a specified point source}

The final output of SW1D, if so desired by the user, is the displacement spectrum at receiver locations specified by their distance $r$ and azimuth $\phi$ from a specified source (epicenter). The source mechanism can be specified in terms of fault strike $\sigma$, dip $\delta$, and rake $\gamma$ (for earthquake sources) or directly through the moment tensor elements of the equivalent double couple. SW1D uses a cartesian North, East, Down coordinate convention to describe fault slip, hence the relations between ($\sigma, \delta, \gamma$) and the cartesian components of the shear-dislocation moment tensor, are the same as in \cite{Aki2002}, Chapter 4. 

At the receiver locations, the radial $(r)$, transverse $(t)$ and vertical $(z)$ components of displacement $(u_r,u_t,u_z)$ are obtained using the following frequency-domain expressions \citep{Mendiguren1977}. 

For the Love wave ($u_t$ only):

\begin{equation}
  u_t(r,\phi,\omega) = A_{t} \left[ M_{xy} \cos{2\phi} + \frac{1}{2} \left(M_{yy} - M_{xx} \right) \sin{2\phi} \right] + B_t \left[ M_{yz} \cos{\phi} - M_{xz} \sin{\phi} \right]\text{,}
  \label{eq:uspec_t}
\end{equation}
where
\begin{align}
\begin{split}
    A_t &= -i\frac{kl_1(z)l_1(d_0)}{8cUI_1^L} \sqrt{\frac{2}{\pi kr}} e^{i(kr + \pi/4)} \\
    B_t &= \frac{l_1(z)l_2(d_0)}{8\mu cUI_1^L} \sqrt{\frac{2}{\pi kr}} e^{i(kr + \pi/4)}
 \end{split}
 \label{eq:excit_t}
\end{align}
are the so-called excitation functions, which depend only on the source depth $d_0$.

For the Rayleigh wave ($u_r$ and $u_z$ only):
\begin{equation}
  u_j(r,\phi,\omega) = A_j \left[ M_{xy} \sin{2\phi} + M_{xx} \frac{1}{2} (1 + \cos{2\phi}) + M_{yy} \frac{1}{2} (1 - \cos{2\phi}) \right] + B_j \left[ M_{yz} \sin{\phi} + M_{xz} \cos{\phi} \right] + C_j M_{zz}
 \label{eq:uspec_zr}
\end{equation}
where the subscript $j$ is either $z$ or $r$, and the corresponding excitation functions are:
\begin{align}
\begin{split}
    A_z &= \frac{kr_2(z) r_1(d_0)}{8cUI_1^R} \sqrt{\frac{2}{\pi kr}} e^{i(kr + \pi/4)} \\
    B_z &= i\frac{r_2(z) r_3(d_0)}{8 \mu cUI_1^R} \sqrt{\frac{2}{\pi kr}} e^{i(kr + \pi/4)} \\
    C_z &= \frac{r_2(z)}{8cUI_1^R} \left( \frac{r_4(d_0) - k \lambda r_1(d_0)}{\lambda + 2\mu} \right)\sqrt{\frac{2}{\pi kr}} e^{i(kr + \pi/4)}
\end{split}
\label{eq:excit_z}
\end{align}
and
\begin{align}
\begin{split}
    A_r &= -i\frac{kr_1(z) r_1(d_0)}{8cUI_1^R} \sqrt{\frac{2}{\pi kr}} e^{i(kr + \pi/4)} \\
    B_r &= \frac{r_1(z) r_3(d_0)}{8 \mu cUI_1^R} \sqrt{\frac{2}{\pi kr}} e^{i(kr + \pi/4)} \\
    C_r &= -i\frac{r_1(z)}{8cUI_1^R} \left( \frac{r_4(d_0) - k \lambda r_1(d_0)}{\lambda + 2\mu} \right)\sqrt{\frac{2}{\pi kr}} e^{i(kr + \pi/4)}
\end{split}
\label{eq:excit_r}
\end{align}

We note that ($A_r, B_r, C_r$) differ from ($A_z, B_z, C_z$) only in that $r_2(z)$ is replaced with $-ir_1(z)$. Equations \eqref{eq:uspec_t}-\eqref{eq:excit_r}, written in a slightly different form, were derived by \cite{Mendiguren1977} using the method of \cite{Saito1967}. In Appendix \ref{app:theory_details}, we show how they relate to the equations of \cite{Aki2002}, which make use of the Love and Rayleigh terms of Green's function.


\section{Benchmarking and testing}
\label{sec:testing}

We present a series of benchmarks to demonstrate the accuracy and reliability of SW1D, including numerical verification of known theoretical relations, comparisons with other software packages and comparisons with published results in the seismological literature.

\subsection{Theoretical validation}

The eigenvalues and eigenfunctions of Love and Rayleigh waves satisfy well-known relations involving their energies, which can be expressed in terms of the energy integrals defined in Appendix \ref{app:theory_details}. We explicitly compute the energy integrals in post-processing (through numerical integrations over the depth coordinate $z$) and verify two sets of relations:

\begin{enumerate}
    \item Averaged over a cycle, a surface wave mode contains equal amounts of kinetic and elastic energy. This is expressed by the equations
    \begin{equation}
        \omega^2I^L_1 = k^2I^L_2 + I^L_3\text{,}
        \label{eq:egy_lov}
    \end{equation}
    for Love waves, and
    \begin{equation}
        \omega^2I^R_1 = k^2I^R_2 + kI^R_3 + I^R_4\text{,}
        \label{eq:egy_ray}
    \end{equation}
    for Rayleigh waves \citep[Chapter 7,][]{Aki2002}.
    We verify equations \eqref{eq:egy_lov} and \eqref{eq:egy_ray} by independently evaluating their LHS and RHS. The calculations are performed separately for each mode-period combination.

    \item The flux of energy ($F$) transported by a surface wave mode $=$ energy density ($E$) $\times$ group velocity ($U$). To verify this relation, the RHS is evaluated using the value of $U$ produced by SW1D, and $E$ obtained as the sum of elastic and kinetic energies (using equation \eqref{eq:egy_lov} or \eqref{eq:egy_ray}, as the case may be). We evaluate the LHS of this relation, the energy flux $F$, using orthogonality relations for Love and Rayleigh waves, which were first provided by \cite{Herrera1964}. Simplified for 2-D wave propagation with the $x$-coordinate along the propagation direction, the \cite{Herrera1964} orthogonality product can be written as:



\begin{equation}
\label{eq:sep_herrera}
\frac{1}{2}\int_0^{\infty}\sum_{q=x,y,z}\left[u_q^l\sigma_{xq}^{m*}-u_q^{m*}\sigma_{xq}^l\right]dz = N^2_m \delta_{lm}\text{,}
\end{equation}

\noindent where $u_q$ and $\sigma_{xq}$ are the components of displacement and stress respectively, on a surface normal to the propagation direction, $l$ and $m$ are mode indices and $\delta_{lm}$ is the Kronecker delta. When $l=m$, Herrera's norm $N_m$ is directly related to the energy flux carried by mode $m$. Working in the frequency domain, one can write $F_m = i \omega N^2_m$, because flux is given by stress $\times$ velocity, and the factor of $i\omega$ transforms displacement in \eqref{eq:sep_herrera} to velocity. Omitting the mode index for clarity, the particular form of F is (Appendix x):
\begin{align}
    \label{eq:flux_herrera_lov}
    F &= \omega k\int_0^{\infty}\mu l_1^2\,dz \text{, for Love modes,} \\
    \label{eq:flux_herrera_ray}
    F &= \omega \int_0^{\infty}\left[\sigma_{xx}r_1 - r_2r_3 \right]\,dz \text{, for Rayleigh modes.}
\end{align}

The relation
\begin{equation}
    F = E U
\label{eq:egy_flux_density}
\end{equation}
is easily proved for both Rayleigh and Love wave modes (see Appendix x) by writing $F$ in terms of the appropriate energy integrals and recognizing that
\begin{align}
    \label{eq:egy_density_lov}
    E &= 2\omega^2 I^L_1 \text{, for Love modes,} \\
    \label{eq:egy_density_ray}
    E &= 2\omega^2 I^R_1 \text{, for Rayleigh modes.}
\end{align}

\end{enumerate}

\begin{figure}[h]
    \centering
    \includegraphics[width=\textwidth]{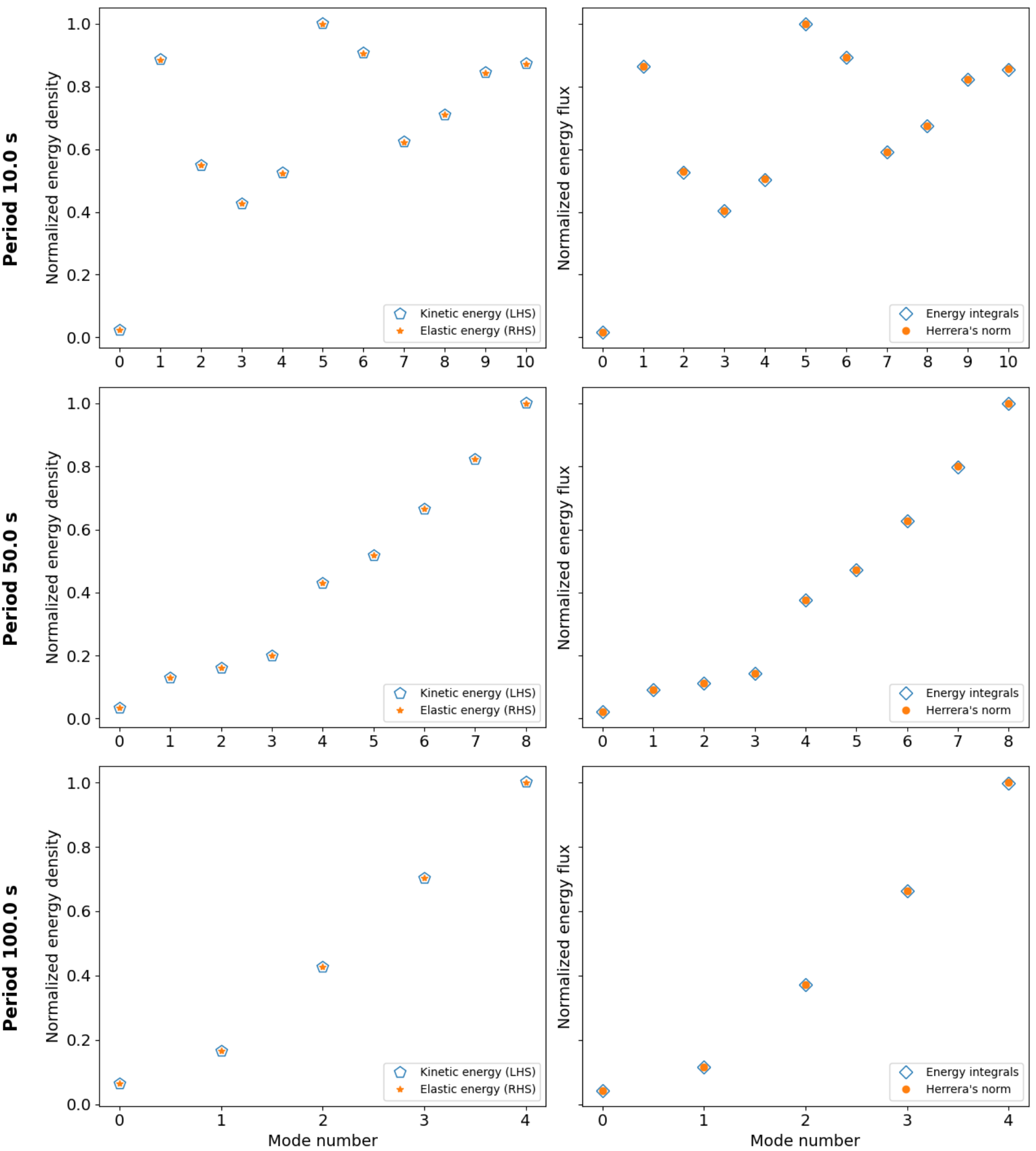}
    \caption{Verification of theoretical relations for Love wave modes. The three rows correspond to periods of 10, 50 and 100 s as labelled on the left-hand side, and values are shown as a function of mode number, with number 0 denoting the fundamental mode. Note the fewer number of modes at the longer periods. In each row, the left column plots the LHS and RHS of equation \eqref{eq:egy_lov}, normalized by the maximum value of the RHS (elastic energy density) across all modes. The right column plots the LHS and RHS of equation \eqref{eq:egy_flux_density} (using equations \eqref{eq:flux_herrera_lov} and \eqref{eq:egy_density_lov}), normalized by the maximum value of the LHS (flux calculated using Herrera's norm) across all modes. The normalizations are done separately for each period.}
    \label{fig:check_egy_int_lov}
\end{figure}

\begin{figure}[h]
    \centering
    \includegraphics[width=\textwidth]{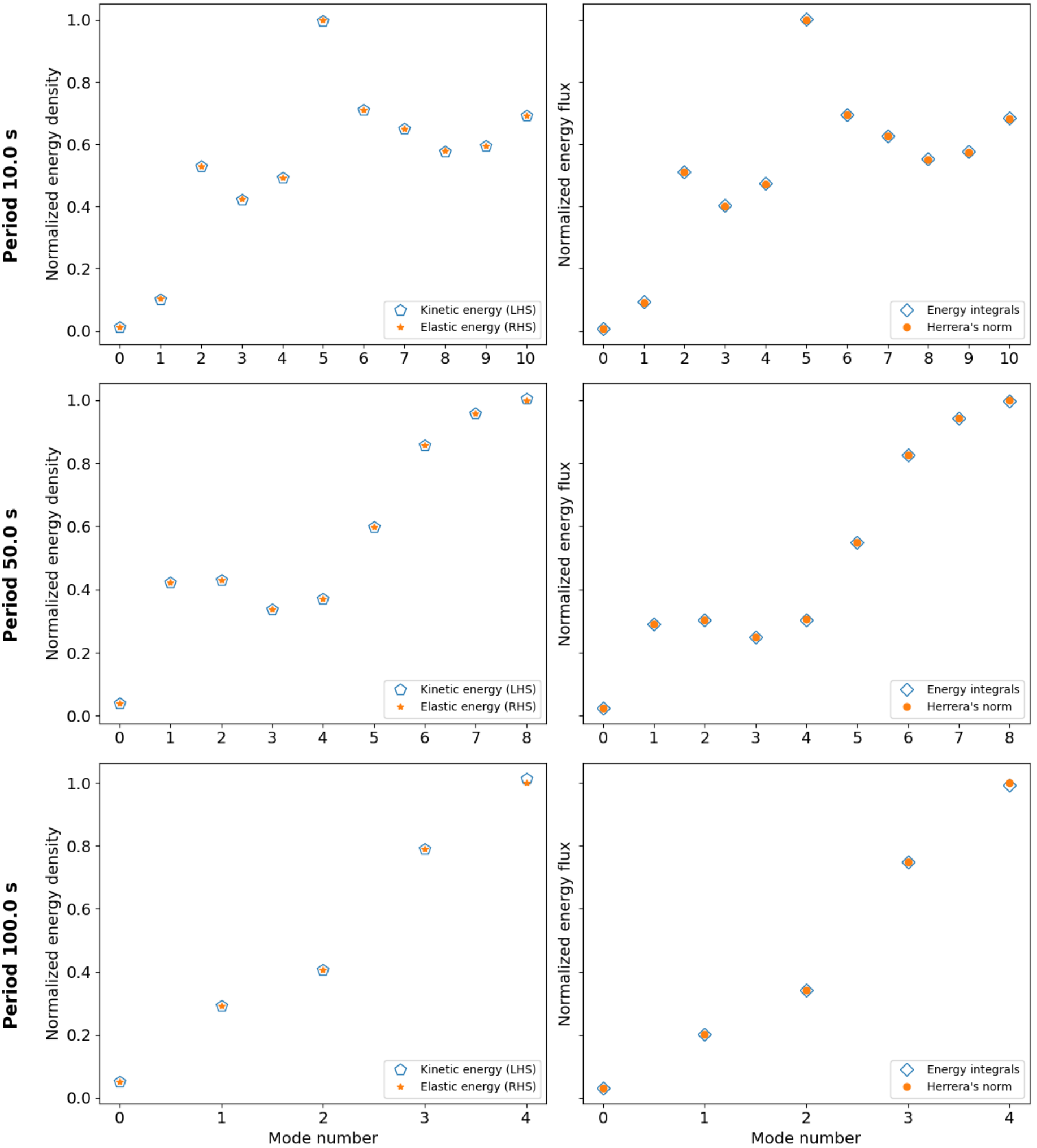}
    \caption{Verification of theoretical relations for Rayleigh wave modes. Similar to Figure \ref{fig:check_egy_int_lov}, except that the plots in the left column correspond to equation \eqref{eq:egy_ray} and those in the right column correspond to equations \eqref{eq:flux_herrera_ray} and \eqref{eq:egy_density_ray} used in \eqref{eq:egy_flux_density}. }
    \label{fig:check_egy_int_ray}
\end{figure}

Using the example of PREM \citep{Dziewonski1981} down to the core-mantle boundary, we demonstrate that SW1D solutions obey the above theoretical relations. The equalities contained in the relations \eqref{eq:egy_lov}-\eqref{eq:egy_ray} and \eqref{eq:flux_herrera_lov}-\eqref{eq:egy_flux_density} are graphed at three different periods, for the fundamental mode and up to ten higher modes, separately for Love and Rayleigh waves (Figures \ref{fig:check_egy_int_lov} and \ref{fig:check_egy_int_ray}, respectively). It is seen that the equalities hold good for nearly all mode-period combinations. Only the highest mode numbers at the longer periods (e.g. 8\textsuperscript{th} higher mode at a period of 50 s or the 4\textsuperscript{th} higher mode at 100 s) exhibit minor discrepancies, which are likely due to incomplete depth-integrals, as the eigenfunctions for these modes extend much deeper than the point to which they are calculated (the bottom of the model, $\approx 2853$ km). Finally, it is noted that whilst Figures \ref{fig:check_egy_int_lov}-\ref{fig:check_egy_int_ray} serve as useful checks on the output of SW1D, they also reflect the quality of numerical integration used to compute the various integrals involved. This in turn depends on the depth sampling used ($\le5$ km in Figures \ref{fig:check_egy_int_lov}-\ref{fig:check_egy_int_ray}), and when the input model is too coarsely sampled, conformity with the theoretical relations can be difficult to establish.

\subsection{Comparison with other software}

\begin{figure}[h]
    \centering
    \includegraphics[width=\textwidth]{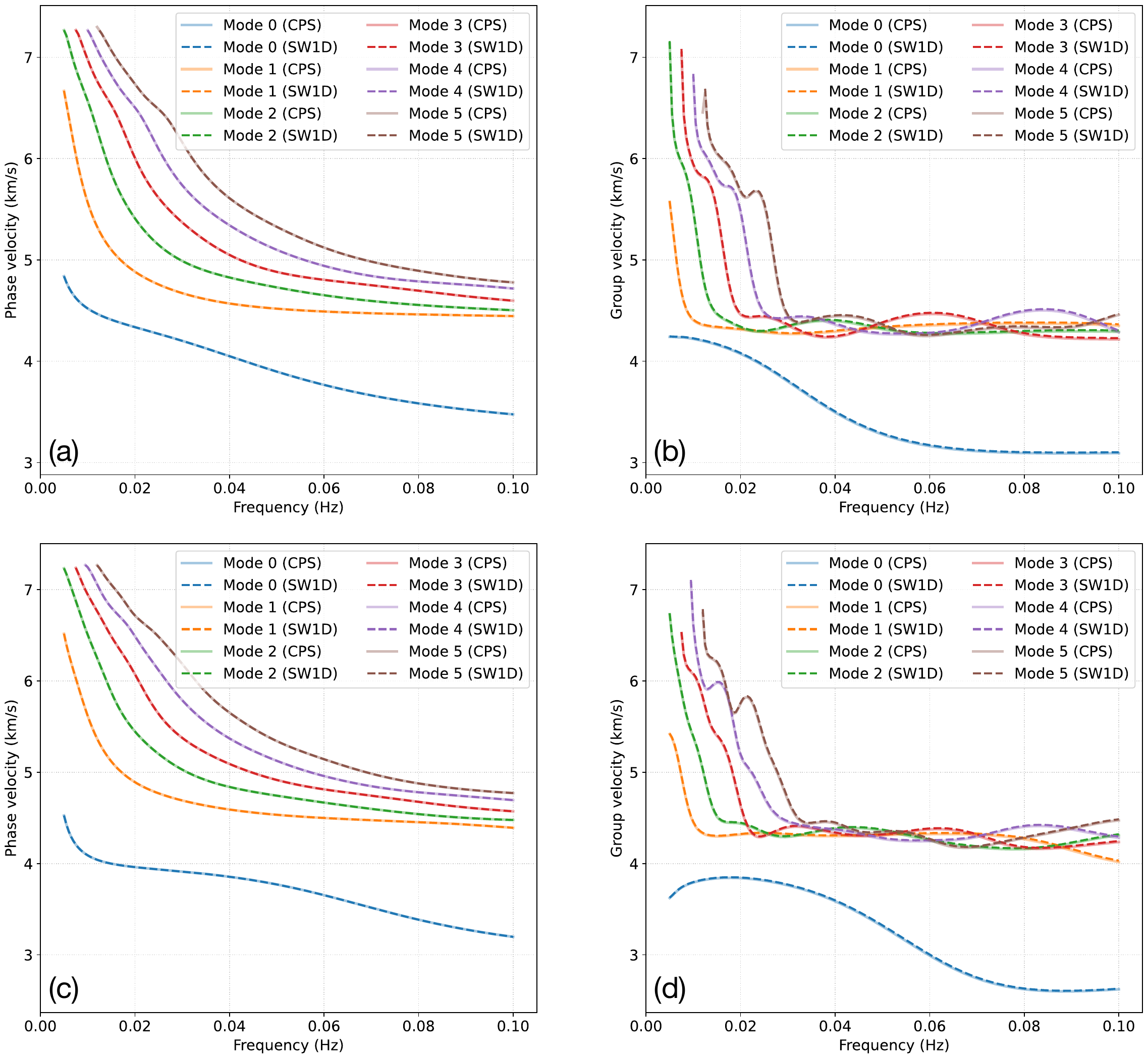}
    \caption{Dispersion curves in PREM, calculated by SW1D (dashed lines) and CPS (solid lines of lighter shade) for the fundamental and first five higher mode surface waves. (a)-(b) Phase and group velocity dispersion respectively, for Love waves. (c)-(d) Phase and group velocity dispersion respectively, for Rayleigh waves. }
    \label{fig:bench_disp}
\end{figure}

\begin{sidewaysfigure}
    \centering
    \includegraphics[width=0.9\textwidth]{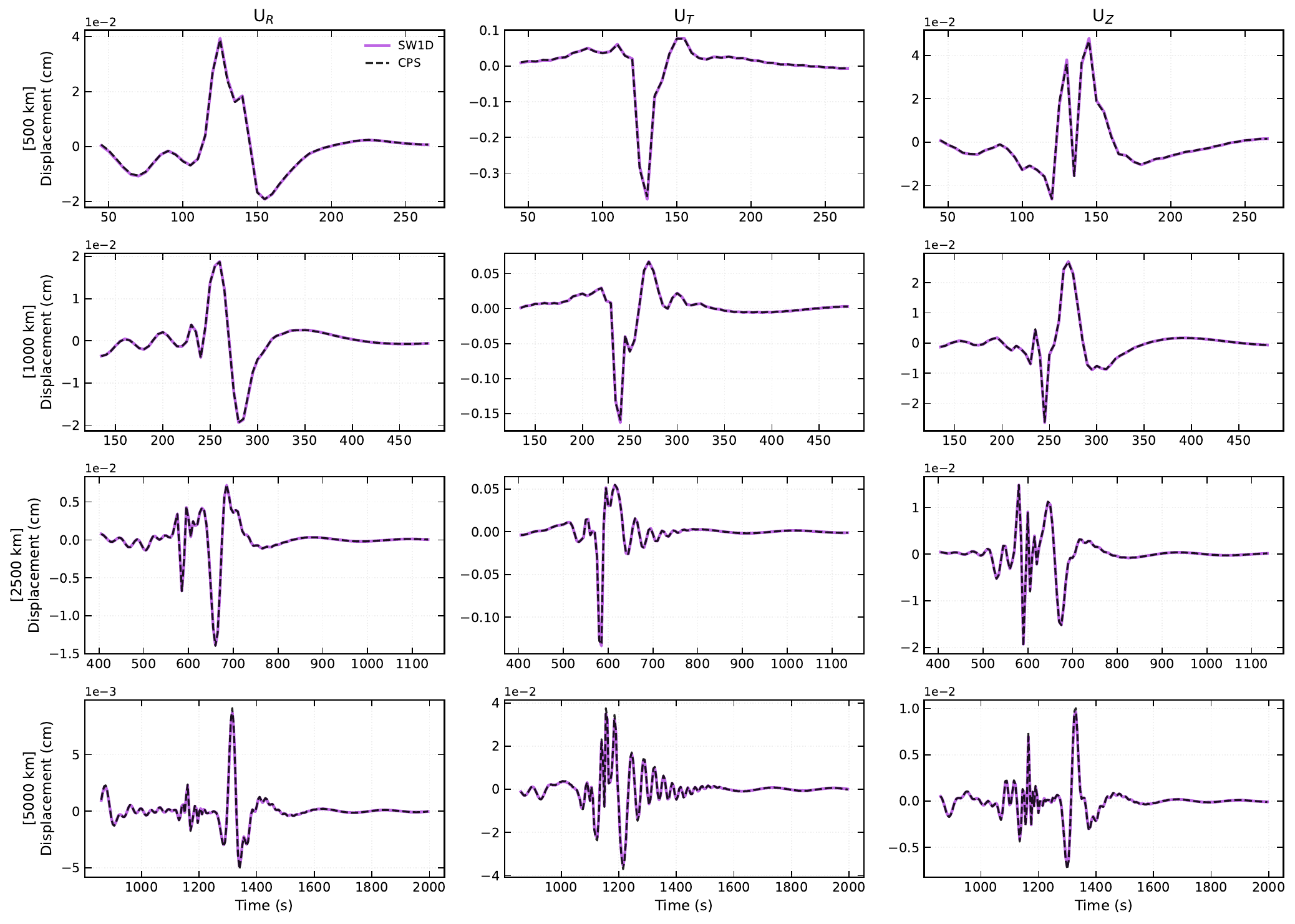}
    \caption{Comparison of synthetic waveforms computed by SW1D (pink solid line) and CPS (black dashed line), for the 152 km-deep, 09 February 2013, $M_w$7.0 Colombia earthquake. The waveforms were generated for a period range of $10-200$ s, with a triangular source time function of half-width $7.5$ s (taken from the GCMT catalog). Shown are the three components of displacement (radial, transverse and vertical, columns from left to right) calculated at different epicentral distances ($500-5000$ km, rows from top to bottom) as labelled. Note the different scales of the individual sub-plots.}
    \label{fig:bench_wforms_col}
\end{sidewaysfigure}

We compare SW1D results with those of the well-known Computer Programs in Seismology software package \citep[CPS;][]{Herrmann2013}, version 3.30. Results are presented for the same Earth model as in the previous section, i.e. PREM underlain by a half-space below a depth of $\approx 2853$ km. Figure \ref{fig:bench_disp} compares phase and group velocity dispersion curves calculated by SW1D and CPS for both Love and Rayleigh waves, for periods longer than $10$ s (frequency < $0.1$ Hz), while Figures \ref{fig:bench_egn_lov}-\ref{fig:bench_egn_ray_uz} in Appendix \ref{app:figs_extra} show the corresponding displacement eigenfunctions at select periods. There is almost no visually discernible difference in the outputs of SW1D and CPS, in any of these figures. Going a step further, Figure \ref{fig:bench_wforms_col} shows synthetic seismograms computed using the centroid moment tensor of a randomly chosen earthquake, as obtained from the Global Centroid-Moment-Tensor (GCMT) catalog \citep{Ekstrom2012}. Again, the waveforms computed by SW1D and CPS are visually indistinguishable on all three displacement components. We note that Earth-flattening transformations have been applied in both SW1D and CPS to produce Figure \ref{fig:bench_wforms_col}, whereas they were not applied in the figures showing the dispersion and eigenfunctions.


\subsection{Comparison with literature}

Numerous seismological studies have been devoted to exploring the relationship between the amplitude spectra of surface waves, and the depth and focal mechanisms of seismic sources such as earthquakes and underground nuclear explosions \citep[e.g.][]{Molnar1969,Tsai1970,Tsai1971,Romanowicz1984}. As an example, we consider the work of \cite{Tsai1971}, who argued that long-period Rayleigh wave spectra could be used to distinguish between explosions and small earthquakes, because of the distinctly different shapes as well as depth-dependence of the Rayleigh wave spectrum, for different types of sources. Their results, which are reproduced in \cite{Aki2002}, showed that earthquake-generated Rayleigh waves exhibit far greater variation with source mechanism and depth, as compared to those generated by explosions.

\begin{figure}[hb]
    \centering
    \includegraphics[width=0.8\textwidth]{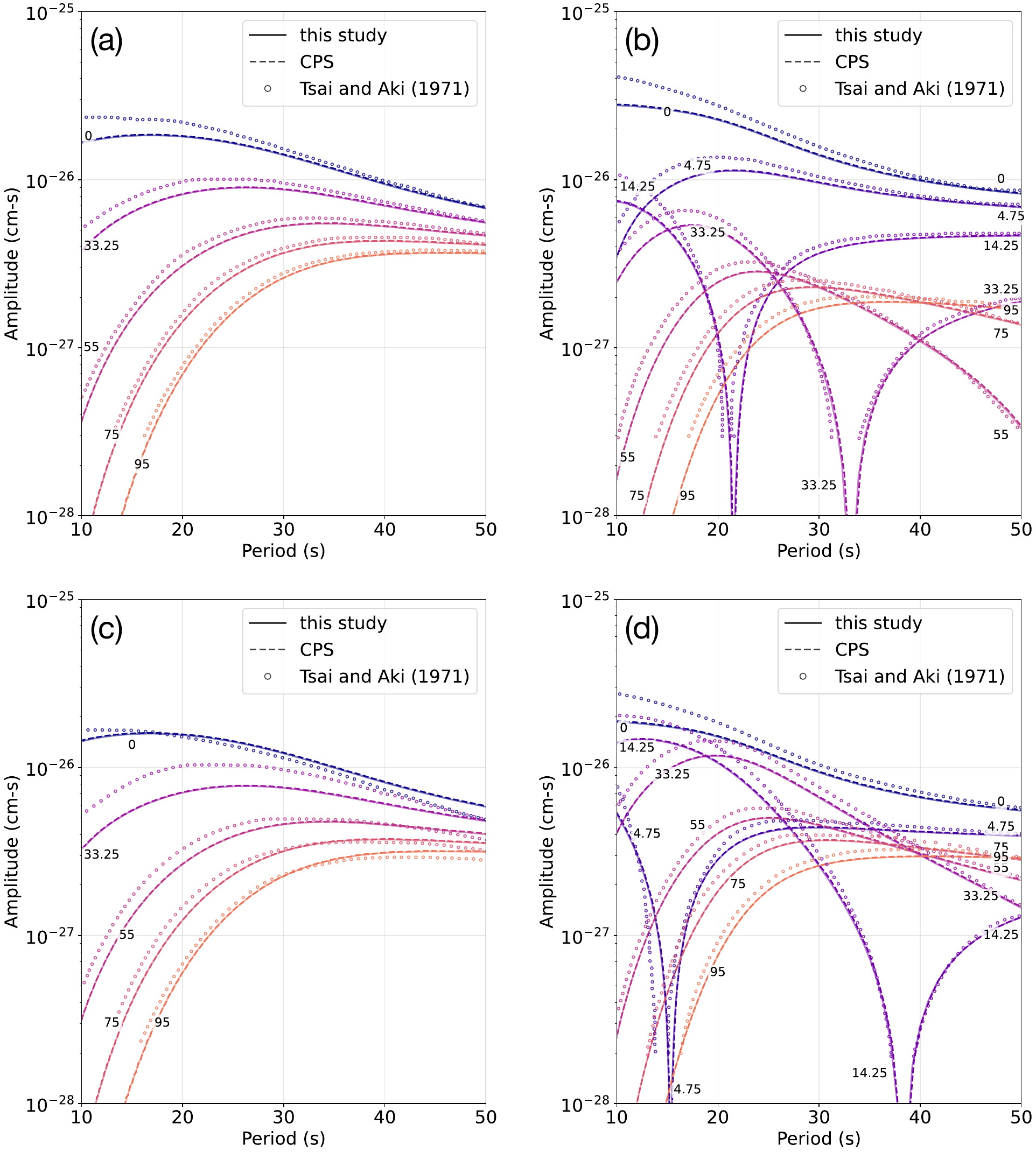}
    \caption{Reproduction of the results of \cite{Tsai1971}. Love wave (left panel) and Rayleigh wave (vertical component, right panel) amplitude spectra for two different earthquake sources: a vertical strike slip earthquake (a,b) and a pure dip-slip earthquake on a fault with dip angle $45^{\circ}$ (c,d). In both cases, a unit step-function seismic moment (1 dyn-cm) is assumed for the source; the receiver is at a distance of 2000 km and azimuth $30^{\circ}$ from the fault strike. In all plots, the numbers beside the curves represent the source depth in km. These plots correspond to Figures 6 and 7 of \cite{Tsai1971}.}
    \label{fig:bench_TsaiAki_6-7}
\end{figure}

\cite{Tsai1971} used the method of \cite{Saito1967} to produce theoretical models of surface wave excitation by explosions and small earthquakes occurring at various depths. In this section, we present the results of undertaking the same theoretical modelling with SW1D. We make sure to use the same Earth model --- the Gutenberg continental Earth model \citep[][Box 7.7]{Aki2002} --- and source mechanisms as were used by \cite{Tsai1971}. Figure \ref{fig:bench_TsaiAki_6-7} compares our results with those of \cite{Tsai1971}, for two types of earthquake sources --- pure strike-slip motion on a vertical fault and pure dip-slip motion on a fault with dip angle $45^{\circ}$. Values from \cite{Tsai1971} are obtained by digitizing the corresponding figures from the electronic version of the paper. For reference, we have also shown the results obtained using CPS. We find a close match between the SW1D and CPS results, but some mismatch with the original results of \cite{Tsai1971}. This could be attributed in part to errors in the digitization. However, the overall character of the original paper's results, i.e.\ the smooth variation of the Love wave spectra and the far more complicated variation of the Rayleigh wave spectra with focal depth, is faithfully reproduced. In Figure \ref{fig:bench_TsaiAki_5}, Appendix \ref{app:figs_extra}, we reproduce the result that for underground explosions, which do not generate Love waves, the Rayleigh wave spectrum varies smoothly with source depth (similar to Love waves for earthquakes) and shows little sensitivity to it in the upper few kilometers.

\clearpage

\section{Applications}
\label{sec:app}

The most fundamental application of SW1D is the forward modelling of surface wave dispersion curves and synthetic seismograms in layered, isotropic, elastic media (as demonstrated in Section \ref{sec:testing}).  However, this is a routine computational seismology task also easily accomplished by other open-source software packages like CPS. On the other hand, we have developed specialized applications of SW1D by integrating it with other computational tools that require fast and reliable surface wave solutions in 1-D Earth models. The end result is niche seismological modelling, insights and understanding. Here we showcase two such unconventional applications of SW1D.



\subsection{Study of surface wave mode conversion}

It is well known from the theory of surface wave propagation in laterally heterogeneous media, that strong lateral variations in the medium produce scattering effects including conversion of surface wave modes among themselves \citep{Kennett1984a,Maupin1987,Odom1986,Snieder2002}. This has implications for several aspects of seismology, such as the validity of the path-average approximation in surface wave tomography \citep{Datta2017}, the amplification of surface waves due to edge effects in sedimentary basins \citep{Brissaud2020}, or the propagation of surface waves across ocean-continent margins \citep{Gregersen1976,Meier2000}. The vast literature on this last topic \citep[e.g.][]{Alsop1966,Gregersen1974,Its1985,Meier1997a} led to the development of the SWRT code by \cite{Datta2018}, to model the interaction of surface waves with a plane vertical discontinuity. Here we revisit the work of \cite{Datta2018}, by combining SWRT with SW1D.

\begin{figure}[b]
    \centering
    \includegraphics[width=\textwidth]{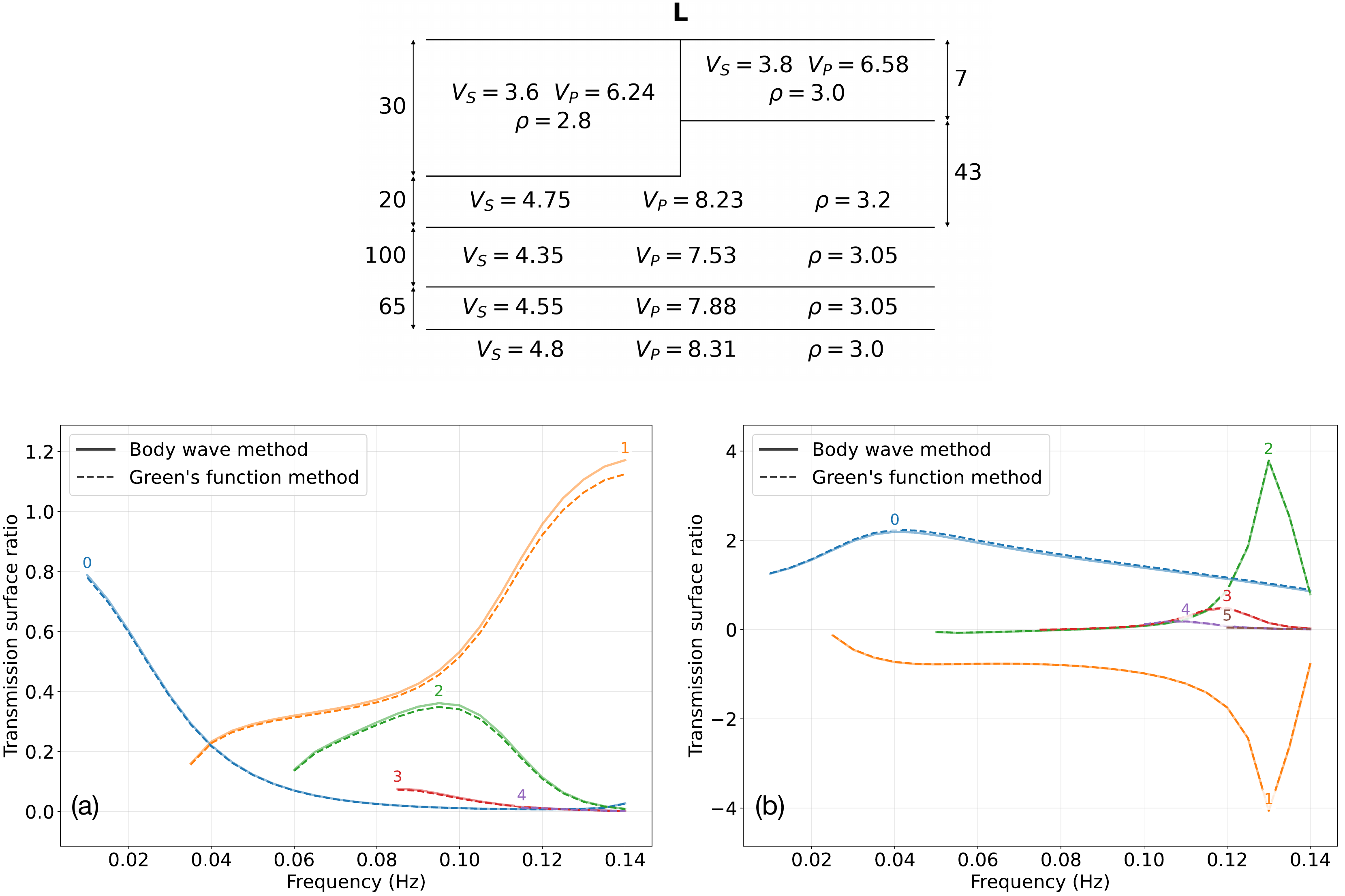}
    \caption{Top: schematic representing `model L'. Layer thicknesses are shown in km, seismic wavespeeds and density are in units of km s\textsuperscript{-1} and gm cc\textsuperscript{-1} respectively. Bottom: Transmission coefficients, with regard to surface displacement, of an incident fundamental mode Love wave for wave propagation in the (a) `forward' direction, i.e.\ from left to right and (b) `backward' direction, i.e.\ from right to left in the model. In each case there are two sets of curves, corresponding to two methods independently implemented in SWRT --- the body wave method of \cite{Gregersen1974} and the Green's functions method of \cite{Its1985}. The numbers beside the curves indicate the mode branch, with 0 denoting the fundamental mode. Note the extent of mode conversion for wave propagation in either direction, and the significant wave amplification (transmission surface ratio $>1$) for propagation in the backward direction.}
    \label{fig:app_SWRT_modL_Love}
\end{figure}

One set of results from the SWRT-SW1D combination, is presented in Figure \ref{fig:app_SWRT_modL_Love}. It shows that a fundamental mode Love wave undergoes multiple mode conversions as it crosses the vertical discontinuity in the popular `model L', which was used by \cite{Gregersen1974} as well as \cite{Its1985}. Figure \ref{fig:app_SWRT_verif} in Appendix \ref{app:figs_extra}, which includes a companion `model F' for completeness, demonstrates good agreement between SWRT results and the original literature, for fundamental mode transmission. We note that the original literature did not show conversions to higher modes for the particular example of Love waves in model L. However, SWRT calculations, including mode conversions, have been verified numerically using finite difference simulations \citep{Datta2016,Datta2018}. SWRT-predicted mode conversion of Rayleigh waves was also used to explain basin edge-induced amplification in \cite{Brissaud2020}.

\subsection{Multi-component ambient noise cross-correlation modelling}

Seismic ambient noise cross-correlations can be modelled under the assumption of uncorrelated sources, for arbitrary spatial distributions of the noise sources \citep{Tromp2010,Fichtner2017,nakata_theoretical_2019}. When this modelling is performed under the elastic regime, it is possible to generate all 9 components of the cross-correlation tensor \citep{Sager2018,Xu2019a,Malkoti2021}. Multi-component noise cross-correlations have seen wide-ranging applications such as measurement of Rayleigh wave ellipticity \citep{Lin2014}, identification of higher mode Rayleigh waves \citep{Ma2016,Nayak2020}, the study of seismic velocity changes caused by earthquakes \citep{Hobiger2014}, inversion for noise source strength distributions \citep{Xu2019a,Xu2020} and adjoint tomography \citep{Wang2019b}.

\begin{figure}[hb]
    \centering
    \includegraphics[width=\textwidth]{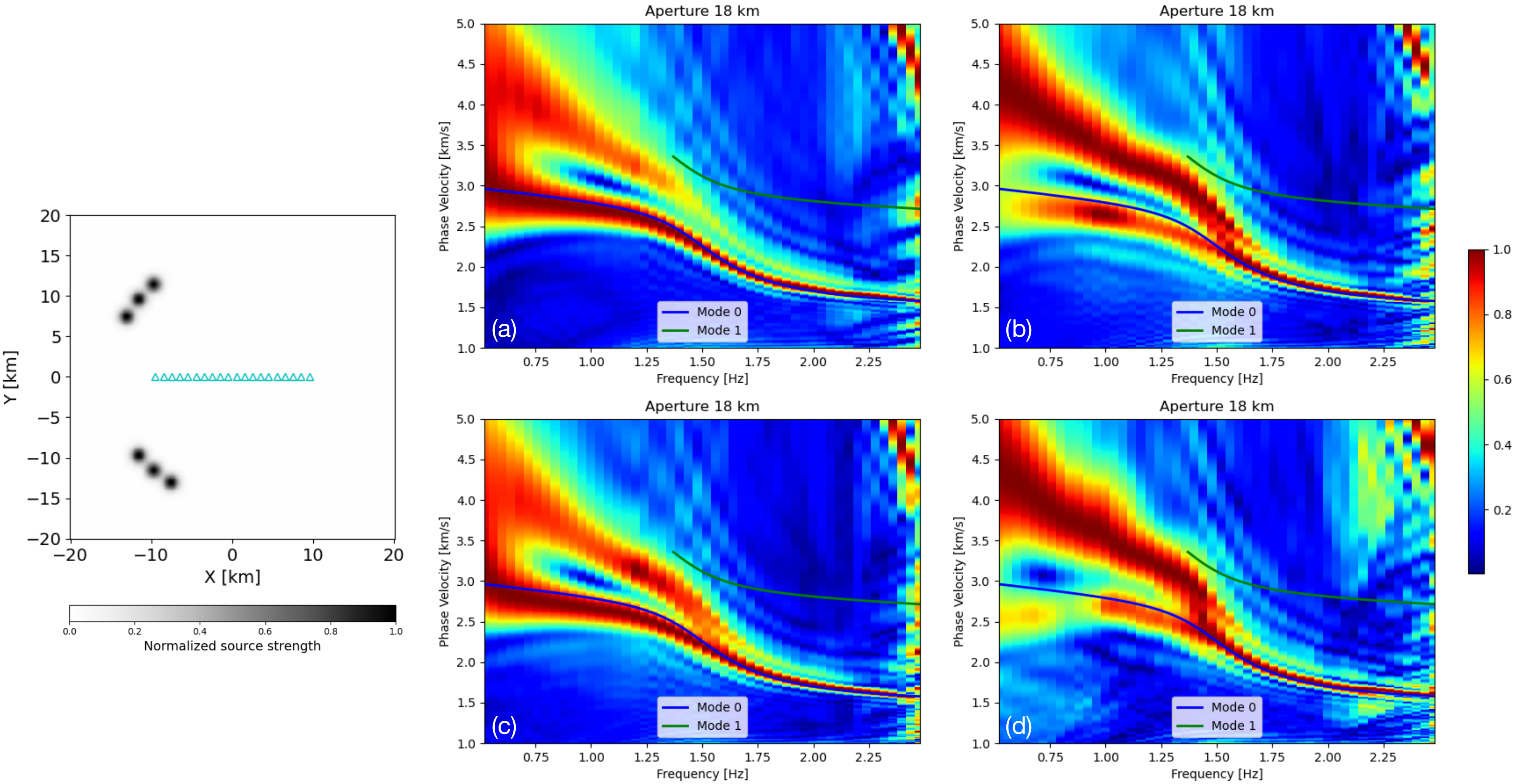}
    \caption{Left: computational domain for the numerical experiment showing the ambient noise source distribution (greyscale) and a linear profile of twenty receivers (triangles); Right: phase velocity dispersion recovered from the waveforms of forward modelled cross-correlations $C_{RR}$ (a), $C_{RZ}$ (b), $C_{ZR}$ (c) and $C_{ZZ}$ (d).  The $f-c$ spectra in (a)-(d) are determined using the slant-stack method as implemented by \cite{Datta2019}. The solid lines on each plot represent the theoretical Rayleigh wave phase velocities in the medium for the fundamental mode and first overtone, as calculated by SW1D. }
    \label{fig:app_mccc_src_DML26}
\end{figure}

Recently, \cite{Datta2026FWANSI2D} presented a package, FW-ANSI2D, that demonstrated elastic modelling of ambient noise cross-correlations using suitably-modified surface wave terms of Green's tensor for a 1-D elastic medium \citep{Malkoti2021,Datta2023b}. Here we integrate FW-ANSI2D with SW1D and repeat the elastic forward modelling experiment of \cite{Datta2026FWANSI2D}, with the difference that we compute all components of the cross-correlation tensor that are relevant for Rayleigh waves: $C_{ZZ}$, $C_{RR}$, $C_{ZR}$ and $C_{RZ}$. For details of this numerical experiment, including the elastic model used, the reader is referred to Section 5.2 of \cite{Datta2026FWANSI2D}. Here we note that only the fundamental mode was included in the modelling. Figure \ref{fig:app_mccc_src_DML26} shows the results obtained with SW1D, in terms of the phase velocity dispersion recovered from the four different sets of cross-correlations. Strong biases or artefacts, which can be misidentified as higher modes, are seen in the recovered phase velocity, especially from $C_{ZZ}$ waveforms. These artefacts can be attributed to the strongly anisotropic noise source illumination. Our results are in agreement with the work of \cite{Xu2017}, who showed that $C_{RR}$ and $C_{ZR}$ are more accurate than $C_{ZZ}$, in the face of strongly inhomogeneous noise source distributions. For reference, Figure \ref{fig:app_mccc_src_iso} in Appendix \ref{app:figs_extra} shows that phase velocities are recovered accurately from all four cross-correlation components, when the source illumination is isotropic around the receivers.

\section{Discussion}

We have presented a Python code, SW1D,  for complete surface wave solutions in 1-D media. Although other codes with this functionality exist and are widely used in the seismological community, our open-source code can be easily integrated with other codes in the Python environment, thereby opening up myriad possibilities for interesting seismological applications (such as those showcased in Section \ref{sec:app}). SW1D is available on GitHub: ttps://github.com/pkolh/SW1D.git.

\clearpage


\appendix

\section{Extended theory}
\label{app:theory_details}

\subsection{Surface wave stress-displacement vector and energy integrals}

Using the notation of \cite{Aki2002}, the stress-displacement vector for Love waves $(l_1,l_2)$ satisfies:

\begin{equation}
\label{eq:sdegn_lov}
    \frac{d}{dz} \begin{pmatrix} l_1 \\ l_2 \end{pmatrix} = 
    \begin{pmatrix} 0 & 1/\mu \\ \mu k^2 - \rho \omega^2 & 0 \end{pmatrix}
    \begin{pmatrix} l_1 \\ l_2 \end{pmatrix}
\end{equation}

while the stress-displacement vector for Rayleigh waves $(r_1, r_2, r_3, r_4)$ satisfies:

\begin{equation}
\label{eq:sdegn_ray}
    \frac{d}{dz} \begin{pmatrix} r_1 \\ r_2 \\ r_3 \\ r_4 \end{pmatrix} = 
    \begin{pmatrix}
    0 & k & \frac{1}{\mu} & 0 \\
    -\frac{k\lambda}{\lambda+2\mu} & 0 & 0 & \frac{1}{\lambda+2\mu} \\
    4k^2\mu \left( \frac{\lambda+\mu}{\lambda+2\mu} \right) - \rho\omega^2 & 0 & 0 & \frac{k\lambda}{\lambda+2\mu} \\
    0 & -\rho\omega^2 & -k & 0
    \end{pmatrix}
    \begin{pmatrix} r_1 \\ r_2 \\ r_3 \\ r_4 \end{pmatrix}
\end{equation}

We note that equations \eqref{eq:sdegn_lov}-\eqref{eq:sdegn_ray} are valid when the elastic parameters $\rho, \lambda, \mu$ are functions of $z$ (depth) only. In the interest of brevity, we have omitted this functional dependence in the above equations.

Energy integrals for Love and Rayleigh waves are defined as follows.

For Love waves:
\begin{align}
\begin{split}
    I_1^L = \frac{1}{2} \int_0^\infty \rho l_1^2 dz\\
    I_2^L = \frac{1}{2} \int_0^\infty \mu l_1^2 dz
\end{split}
\label{eq:egy_int_lov}
\end{align}

For Rayleigh waves:
\begin{align}
\begin{split}
    I_1^R = \frac{1}{2} \int_0^\infty \rho (r_1^2 + r_2^2) dz\\
    I_2^R = \frac{1}{2} \int_0^\infty [(\lambda + 2\mu)r_1^2 + \mu r_2^2] dz\\
    I_3^R = \int_o^\infty \left( \lambda r_1 \frac{dr_2}{dz} - \mu r_2 \frac{dr_1}{dz} \right) dz
\end{split}
\label{eq:egy_int_ray}
\end{align}

\subsection{Surface wave excitations from double-couple sources}

From equation (7.149) of \cite{Aki2002},
\begin{align*}
    u^{\text{L}}_t &= \sum_n \frac{l_1(z)}{8cUI_1^L} \sqrt{\frac{2}{\pi kr}} \exp\left[ i\left( k_n r + \frac{\pi}{4} \right) \right] \times \\
    &\bigg\{ i k_n l_1(d_0) \left( M_{xx} \sin{\phi} \cos{\phi} - M_{yx} \cos^2{\phi} + M_{xy} \sin^2{\phi} - M_{yy} \sin{\phi} \cos{\phi} \right) \\
    &\qquad - \frac{dl_1}{dz} \bigg|_{d_0} \left( M_{xz} \sin{\phi} - M_{yz} \cos{\phi} \right) \bigg\}
\end{align*}

In the above equation, use equation \eqref{eq:sdegn_lov} to replace $\frac{dl_1}{dz} \big|_{d_0}$ with $l_2(d_0)/\mu$. Now using $M_{yx} = M_{xy}$ (since $\mathbf{M}$ is symmetric) and the definitions \eqref{eq:excit_t}, one obtains equation \eqref{eq:uspec_t}.


We now refer to equation (7.150) from \cite{Aki2002}.
\begin{align*}
    u^{\text{R}}_z &= \sum_n \frac{r_2(z)}{8cUI_1^R} \sqrt{\frac{2}{\pi k_n r}} \exp\left[ i \left( k_n r + \frac{\pi}{4} \right) \right] \times \\
    &\qquad \bigg\{ k_n r_1(d_0) \left( M_{xx} \cos^2{\phi} + (M_{xy} + M_{yx}) \sin{\phi} \cos{\phi} + M_{yy} \sin^2{\phi} \right) \\
    &\qquad + i \frac{dr_1}{dz} \bigg|_{d_0} \left( M_{xz} \cos{\phi} + M_{yz} \sin{\phi} \right) - ik_n r_2(d_0) \left( M_{zx} \cos{\phi} + M_{zy} \sin{\phi} \right) + \frac{dr_2}{dz} \bigg|_{d_0} M_{zz} \bigg\}
\end{align*}

Here, use equation \eqref{eq:sdegn_ray} to replace $\frac{dr_2}{dz} \big|_{d_0}$ with $[\lambda + 2\mu]^{-1} \left(r_4(d_0) - k\lambda r_1(d_0)\right)$. Using the symmetry of $\mathbf{M}$ and again using equation \eqref{eq:sdegn_ray} to replace $(\frac{dr_1}{dz} \big|_{d_0} - kr_2(d_0))$ with $r_3(d_0)/\mu$, one obtains equation \eqref{eq:uspec_zr} with the definitions \eqref{eq:excit_z}. Similarly for the radial component, equation (7.151) of \cite{Aki2002} can be manipulated to yield equation \eqref{eq:uspec_zr} with the definitions \eqref{eq:excit_r}.




\subsection{Energy flux relations for surface waves}

\subsubsection{Love waves}

In this case, $u_x=u_z=0$, $\sigma_{xx}=\sigma_{xz}=0$, $u_y \propto l_1(z)$ and $\sigma_{xy}=\mu\frac{\partial u_y}{\partial x}\propto ik\mu l_1(z)$. $l_1(z)$ is the displacement eigenfunction. Substituting in \eqref{eq:sep_herrera}, Hererra's norm (for mode $m$) becomes
\begin{align*}
\label{eq:sep_herr_lov}
N_m&=\frac{1}{2}\int_0^{\infty}\left[l_1(-ik_m\mu l_1) - l_1(ik_m\mu l_1)\right]dz \\
&=-ik_m\int_0^{\infty}\mu(z) {l_1(z)}^2dz
\end{align*}

\noindent Energy flux $F$ = $i\omega \times N$, so
\begin{equation}
\label{eq:sep_flux_lov}
F=\omega k\int_0^{\infty}\mu l_1^2dz
\end{equation}

\noindent It is trivial to see that $F$ is related to the energy density integral $I^L_2$ as $F=2\omega kI^L_2$. Hence the ratio of energy flux to total energy density $E$ is

\begin{equation*}
\frac{F}{E}=\frac{2\omega k I^L_2}{2{\omega}^2I^L_1}=\frac{I^L_2}{cI^L_1}=U\text{,}
\end{equation*}

\noindent where $c$ is phase velocity and $U$ is group velocity.

\subsubsection{Rayleigh waves}

In this case, $u_y=0$ and the relevant components of stress for Eq \eqref{eq:sep_herrera} are $\sigma_{xx}$ and $\sigma_{xz}$. In terms of eigenfunctions and in vector notation, \begin{align*}
\vec{u}&=r_1\hat{x}+ir_2\hat{z} \\
\vec{\sigma_x}&=i\sigma_{xx}\hat{x}+r_3\hat{z}
\end{align*}

\noindent Therefore Herrera's norm \eqref{eq:sep_herrera} can be written (with the mode index $m$ omitted for clarity) as follows:
\begin{align}
\label{eq:sep_herr_ray}
N&=\frac{1}{2}\int_0^{\infty}\left[(r_1\hat{x}+ir_2\hat{z})(-i\sigma_{xx}\hat{x}+r_3\hat{z}) - (r_1\hat{x}-ir_2\hat{z})(i\sigma_{xx}\hat{x}+r_3\hat{z})\right]dz \nonumber \\
&=\frac{1}{2}\int_0^{\infty}\left[-i\sigma_{xx}r_1 + ir_2r_3 - i\sigma_{xx}r_1 + ir_2r_3\right]dz \nonumber \\
&=-i\int_0^{\infty}\left[\sigma_{xx}r_1 - r_2r_3\right]dz
\end{align}

\noindent Therefore energy flux is

\begin{equation}
\label{eq:sep_flux_ray}
F=\omega \int_0^{\infty}\left[\sigma_{xx}r_1 - r_2r_3 \right]dz
\end{equation}

\noindent Writing $\sigma_{xx}$ and $r_3$ in terms of $r_1$ and $r_2$, the above equation can be expressed in terms of the Rayleigh wave energy integrals $I^R_2$ and $I^R_3$: $F=
\omega(2kI^R_2+I^R_3)$. Using $E=2{\omega}^2I^R_1$,

\begin{equation*}
\frac{F}{E}=\frac{\omega(2kI^R_2+I^R_3)}{2{\omega}^2I^R_1}=\frac{kI^R_2 + I^R_3/2}{\omega I^R_1}=\frac{I^R_2+I^R_3/2k}{cI^R_1}=U\text{.}
\end{equation*}

\clearpage

\section{Additional figures}
\label{app:figs_extra}

\begin{figure}[h]
    \centering
    \includegraphics[width=\textwidth]{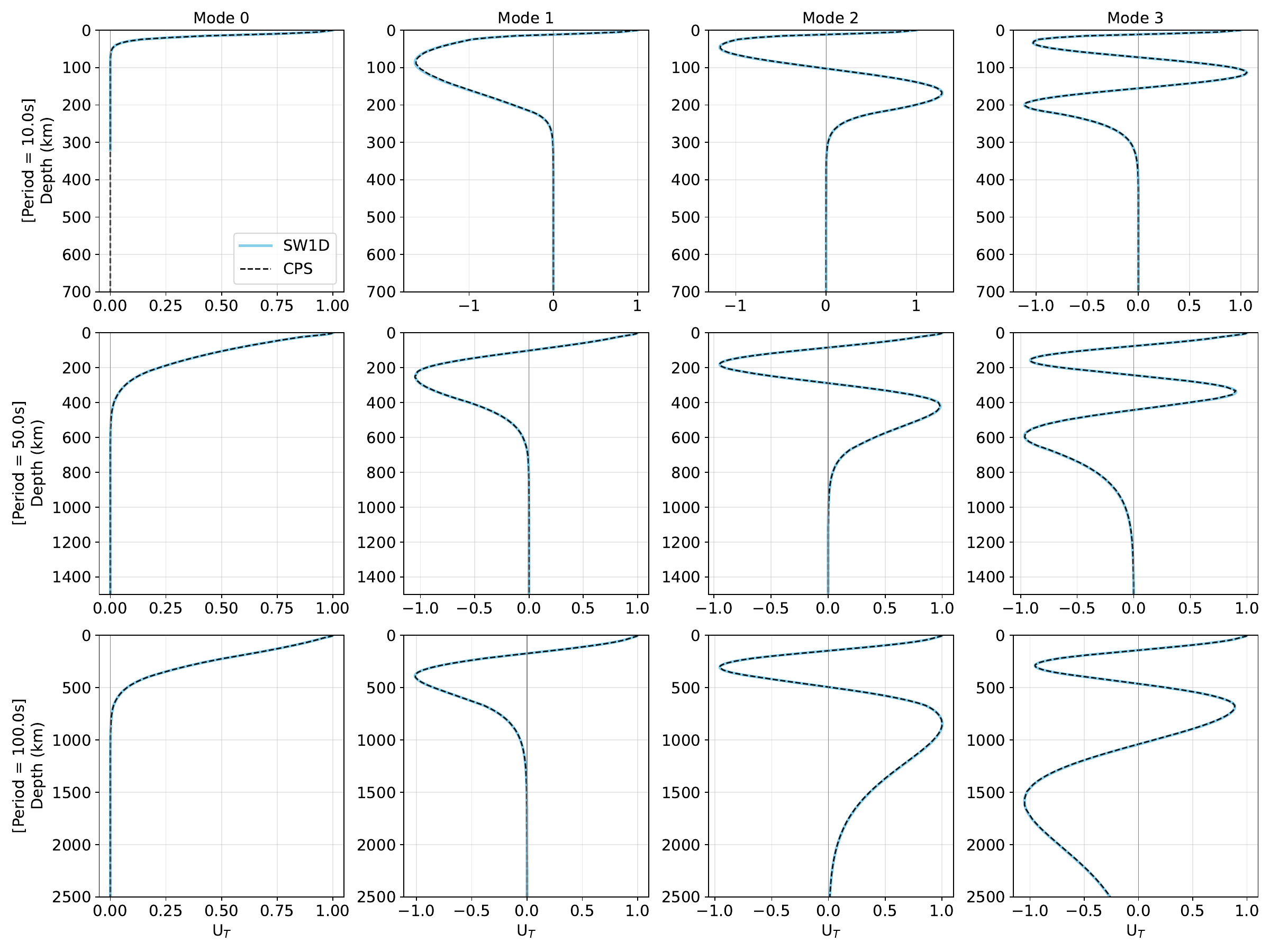}
    \caption{Love wave displacement eigenfunctions computed by SW1D (light blue solid line) and CPS (dark blue dashed line), for different modes (columns) and periods (rows) as labelled at the top and on the left of the figure. The fundamental and first three higher modes are shown at periods of 10 s, 50 s and 100 s. The amplitude is normalized to the displacement at the surface ($z=0$). Note the different $y-$axis limits for different periods.}
    \label{fig:bench_egn_lov}
\end{figure}

\begin{figure}[h]
    \centering
    \includegraphics[width=\textwidth]{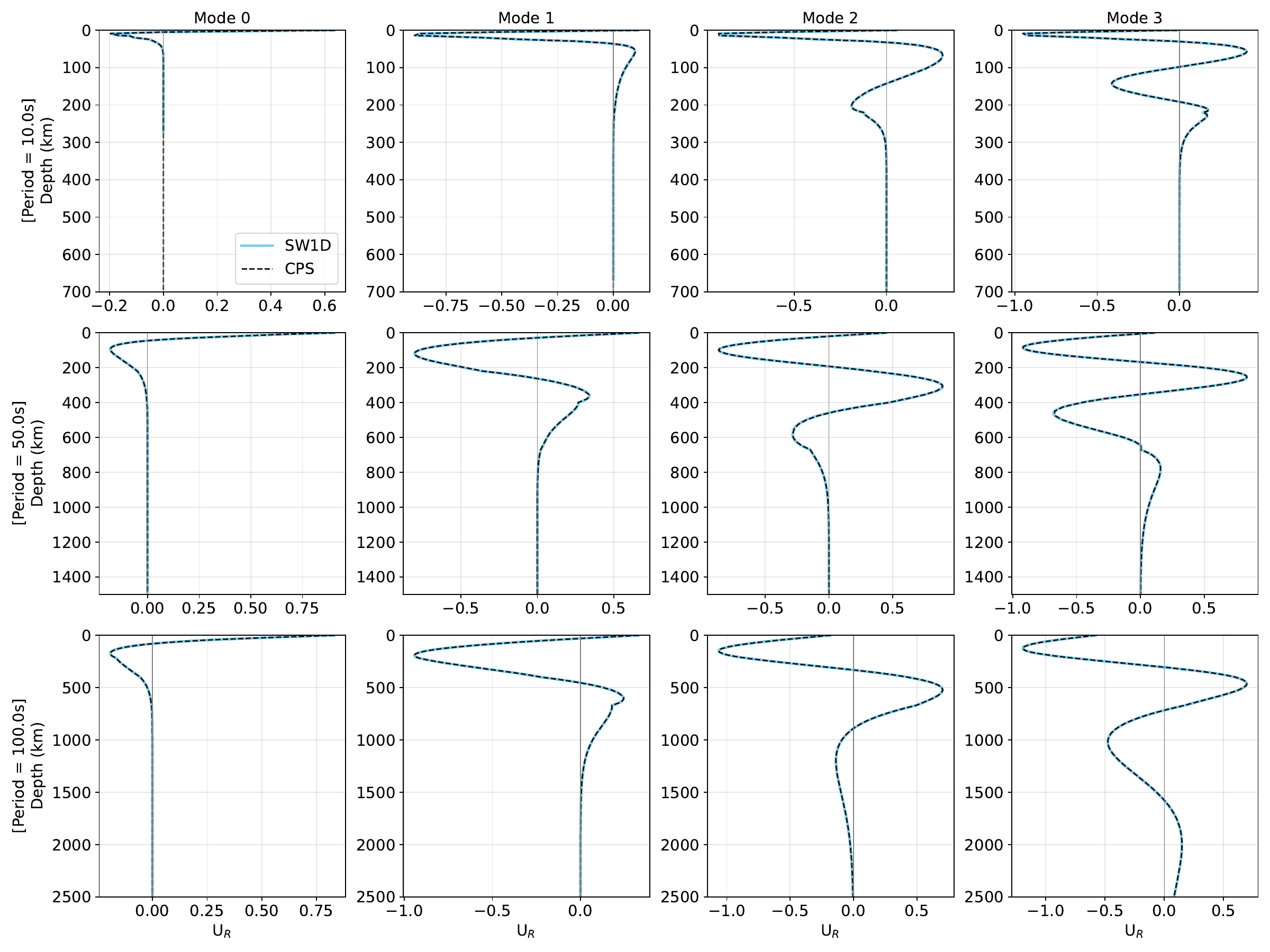}
    \caption{Same as Figure \ref{fig:bench_egn_lov} but for Rayleigh wave displacement eigenfunctions (radial component). The amplitude is normalized to the vertical displacement at $z=0$.}
    \label{fig:bench_egn_ray_ur}
\end{figure}

\begin{figure}[h]
    \centering
    \includegraphics[width=\textwidth]{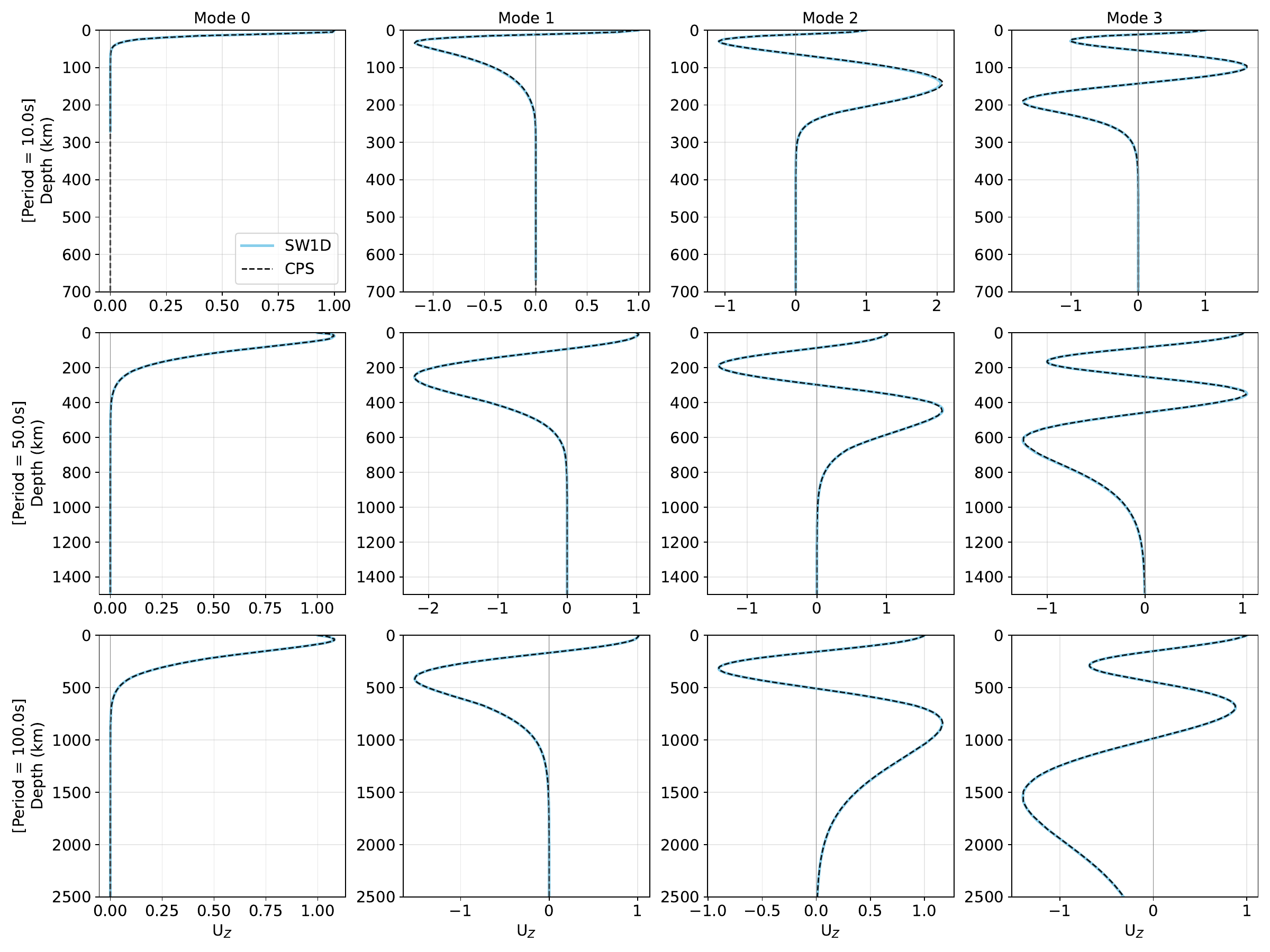}
    \caption{Same as Figure \ref{fig:bench_egn_ray_ur}, but for the vertical component of displacement. }
    \label{fig:bench_egn_ray_uz}
\end{figure}

\begin{figure}[h]
    \centering
    \includegraphics[width=0.5\textwidth]{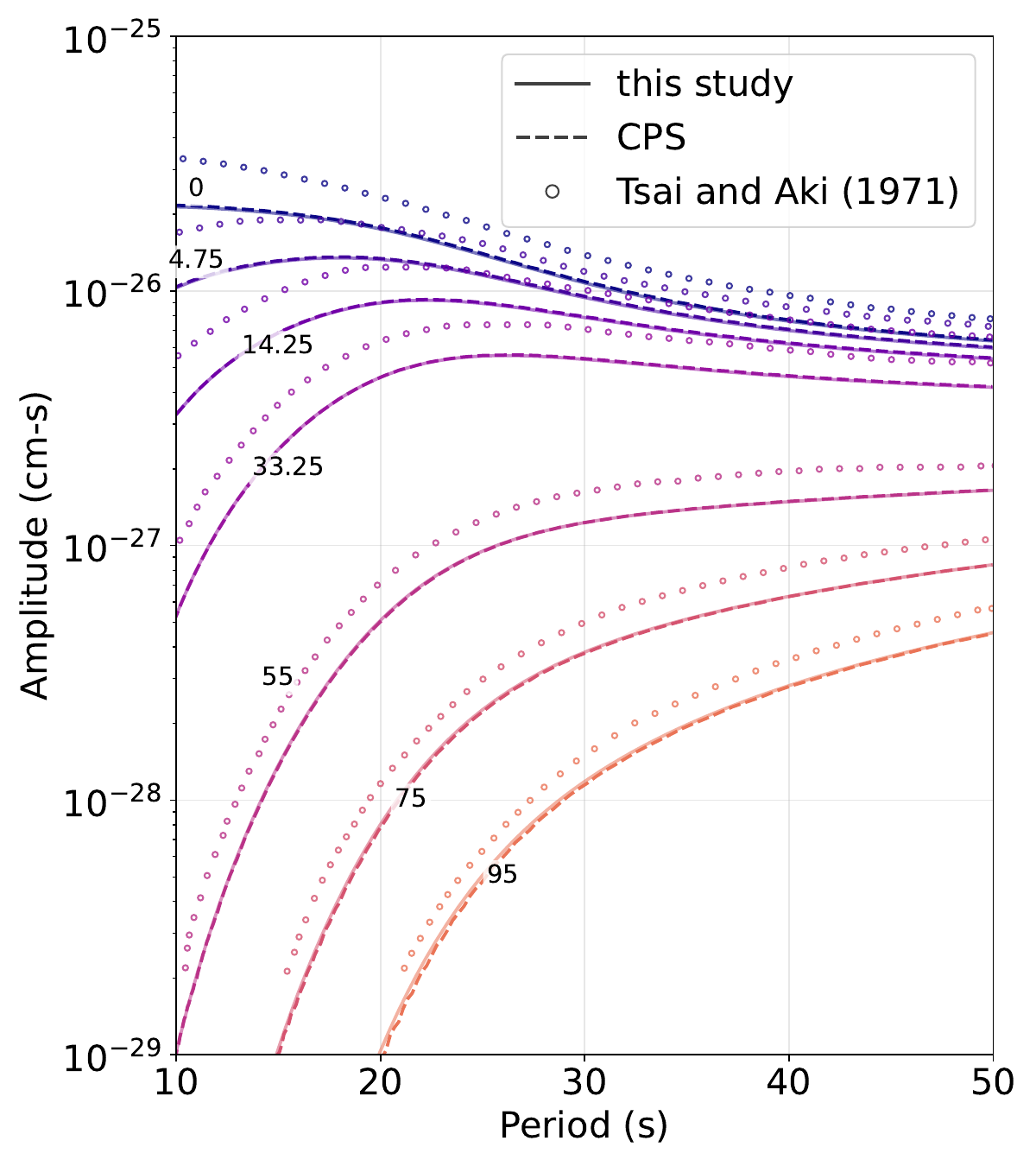}
    \caption{Similar to Figure \ref{fig:bench_TsaiAki_6-7} in the main text, but showing the Rayleigh wave vertical displacement amplitude spectra for underground explosions (with unit seismic moment) at various depths.}
    \label{fig:bench_TsaiAki_5}
\end{figure}

\begin{figure}[h]
    \centering
    \includegraphics[width=\textwidth]{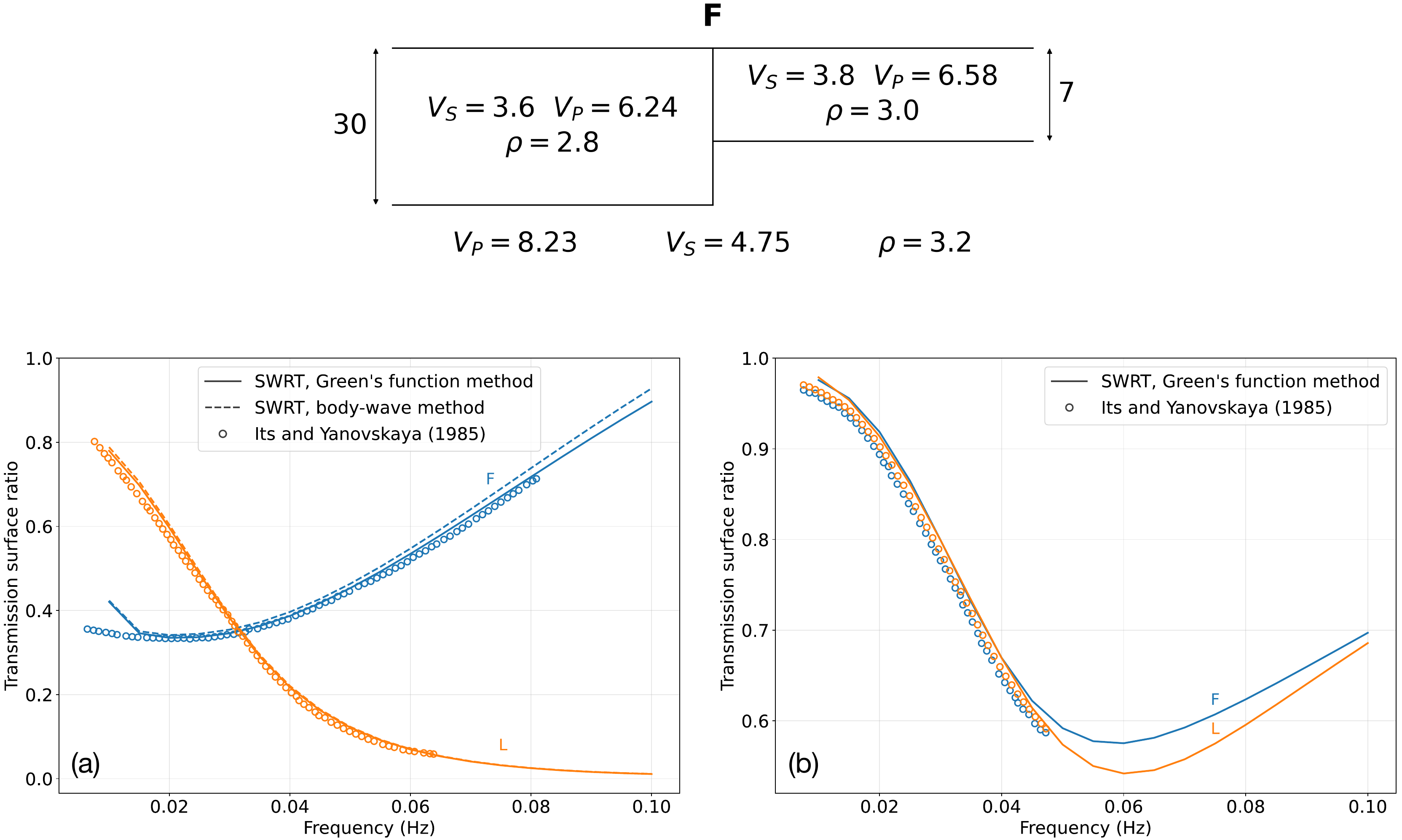}
    \caption{Top: schematic representing `model F', with units of values shown being the same as for `model L' in Figure \ref{fig:app_SWRT_modL_Love}. Bottom: Transmission coefficients, with regard to surface displacement, for forward propagation of a fundamental mode (a) Love wave (b) Rayleigh wave, in models F and L, as indicated by the letters beside the curves. SWRT results (solid or dashed lines) are compared with those of \cite{Its1985} by digitizing figures 5 and 9 of their paper (circles). For a clear comparison, conversion to higher modes in case of the Love wave in model L is not shown, but the fundamental mode result in (a, curve labelled `L') is the same as in Figure \ref{fig:app_SWRT_modL_Love}(a, curve labelled `0'). Note that the body-wave method is not defined for Rayleigh waves and the results of \cite{Its1985} are not available at the high-frequency end of the range considered.}
    \label{fig:app_SWRT_verif}
\end{figure}

\begin{figure}[h]
    \centering
    \includegraphics[width=\textwidth]{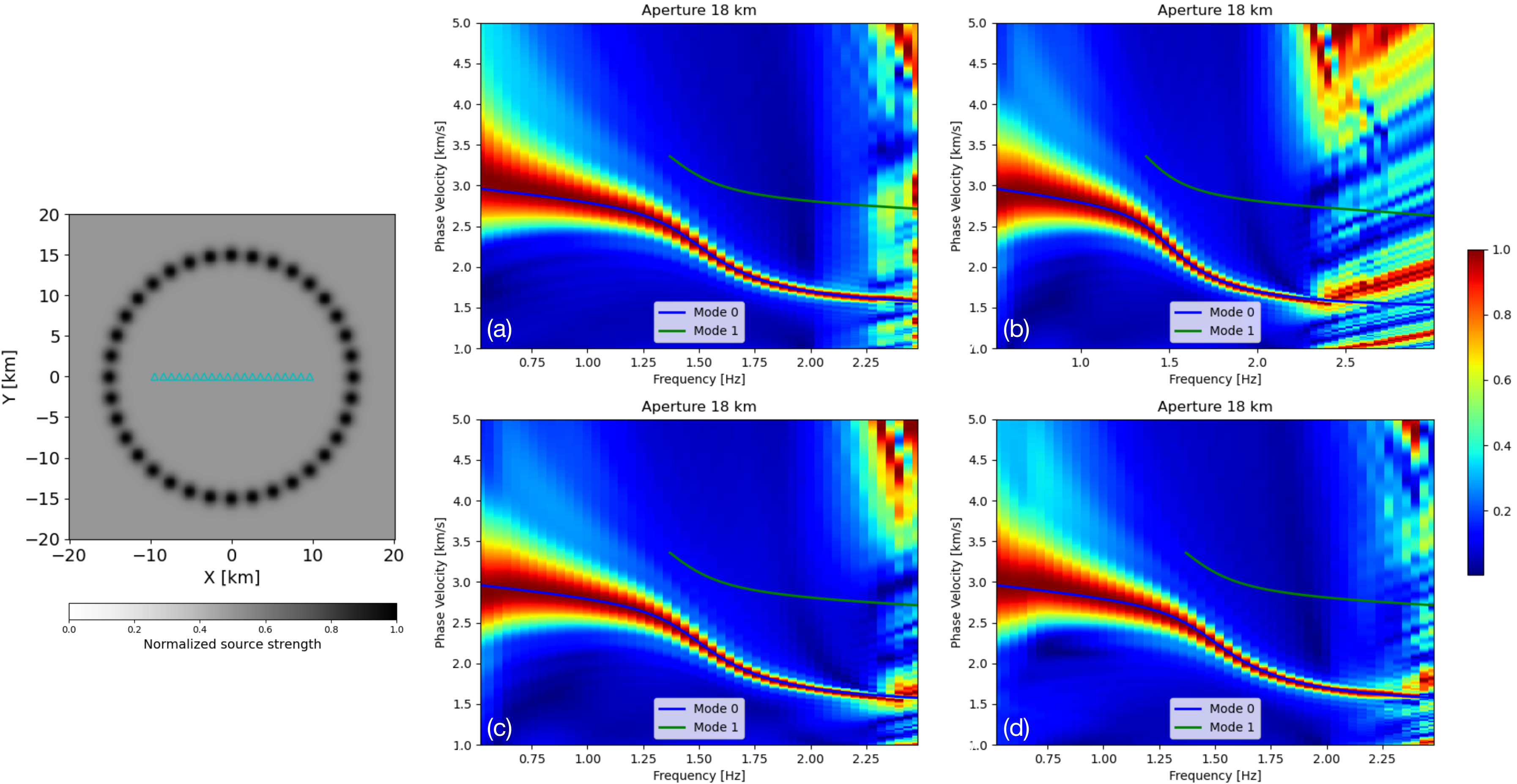}
    \caption{Similar to Figure \ref{fig:app_mccc_src_DML26} of the main text, but for an isotropic source distribution as shown on the left. }
    \label{fig:app_mccc_src_iso}
\end{figure}


\clearpage

\bibliographystyle{plainnat}
\bibliography{ADmendeley}

@incollection{nakata_theoretical_2019,
	edition = {1},
	title = {Theoretical {Foundations} of {Noise} {Interferometry}},
	isbn = {978-1-108-26480-8 978-1-108-41708-2},
	url = {https://www.cambridge.org/core/product/identifier/9781108264808%23c4/type/book_part},
	urldate = {2020-09-21},
	booktitle = {Seismic {Ambient} {Noise}},
	publisher = {Cambridge University Press},
	author = {Fichtner, Andreas and Tsai, Victor C},
	editor = {Nakata, Nori and Gualtieri, Lucia and Fichtner, Andreas},
	month = apr,
	year = {2019},
	doi = {10.1017/9781108264808.006},
	pages = {109--143}
}

@article{Thomson1950,
   author = {William T Thomson},
   issue = {2},
   journal = {J. App. Phys.},
   pages = {89-93},
   publisher = {AIP Publishing},
   title = {Transmission of elastic waves through a stratified solid medium},
   volume = {21},
   year = {1950},
}

@article{Haskell1953,
  author  = {Haskell, N. A.},
  title   = {The Dispersion of Surface Waves on Multilayered Media},
  journal = {Bulletin of the Seismological Society of America},
  year    = {1953},
  volume  = {43},
  number  = {1},
  pages   = {17--34}
}

@book{Pestel1963,
   author = {Eduard C Pestel and Frederick A Leckie},
   publisher = {McGraw-Hill New York},
   title = {Matrix Methods in Elastomechanics},
   volume = {435},
   year = {1963},
}

@article{Herrera1964,
   author = {I Herrera},
   issue = {4},
   journal = {Bull. Seismol. Soc. Am.},
   pages = {1087-1096},
   publisher = {Seismological Society of America},
   title = {On a method to obtain a {G}reen's function for a multi-layered half space},
   volume = {54},
   year = {1964},
}

@article{Knopoff1964matrix,
   author = {L Knopoff},
   issue = {1},
   journal = {Bull. Seismol. Soc. Am.},
   pages = {431-438},
   publisher = {Seismological Society of America},
   title = {A matrix method for elastic wave problems},
   volume = {54},
   year = {1964},
}

@article{Dunkin1965,
   author = {John W Dunkin},
   issue = {2},
   journal = {Bull. Seismol. Soc. Am.},
   pages = {335-358},
   publisher = {Seismological Society of America},
   title = {Computation of modal solutions in layered, elastic media at high frequencies},
   volume = {55},
   year = {1965},
}

@article{Thrower1965,
   author = {Eo No Thrower},
   issue = {3},
   journal = {Journal of Sound and Vibration},
   pages = {210-226},
   publisher = {Elsevier},
   title = {The computation of the dispersion of elastic waves in layered media},
   volume = {2},
   year = {1965},
}

@article{Alsop1966,
   author = {L E Alsop},
   issue = {16},
   journal = {J. Geophys. Res.},
   pages = {3969-3984},
   publisher = {Wiley Online Library},
   title = {Transmission and reflection of Love waves at a vertical discontinuity},
   volume = {71},
   year = {1966},
}

@article{Gilbert1966,
   author = {Freeman Gilbert and George E Backus},
   issue = {2},
   journal = {Geophysics},
   pages = {326-332},
   publisher = {Society of Exploration Geophysicists},
   title = {Propagator matrices in elastic wave and vibration problems},
   volume = {31},
   year = {1966},
}

@article{Biswas1970,
   author = {N N Biswas and L Knopoff},
   issue = {4},
   journal = {Bull. Seismol. Soc. Am.},
   pages = {1123-1137},
   publisher = {Seismological Society of America},
   title = {Exact earth-flattening calculation for Love waves},
   volume = {60},
   year = {1970},
}

@article{Lysmer1970,
   author = {John Lysmer},
   issue = {1},
   journal = {Bull. Seismol. Soc. Am.},
   pages = {89-104},
   publisher = {Seismological Society of America},
   title = {Lumped mass method for {Rayleigh} waves},
   volume = {60},
   year = {1970},
}

@article{Watson1970,
   author = {Watson, T.H.},
   issue = {1},
   journal = {Bull. Seismol. Soc. Am.},
   pages = {161-166},
   publisher = {Seismological Society of America},
   title = {A note on fast computation of Rayleigh wave dispersion in the multilayered elastic half-space},
   volume = {60},
   year = {1970},
}

@article{Schwab1971,
   author = {Fred Schwab and Leon Knopoff},
   issue = {4},
   journal = {Bull. Seismol. Soc. Am.},
   pages = {893-912},
   publisher = {Seismological Society of America},
   title = {Surface waves on multilayered anelastic media},
   volume = {61},
   year = {1971},
}

@article{Biswas1972,
   author = {N N Biswas},
   issue = {1},
   journal = {Pure App. Geophys.},
   pages = {61-74},
   publisher = {Springer},
   title = {Earth-flattening procedure for the propagation of Rayleigh wave},
   volume = {96},
   year = {1972},
}

@incollection{Takeuchi1972,
  author    = {Takeuchi, H. and Saito, M.},
  title     = {{Seismic Surface Waves}},
  booktitle = {Methods in Computational Physics},
  editor    = {Bolt, B. A.},
  publisher = {Academic Press},
  address   = {New York},
  year      = {1972},
  volume    = {11},
  pages     = {217--295}
}

@article{Gregersen1974,
   author = {Søren Gregersen and L E Alsop},
   issue = {3},
   journal = {Bulletin of the Seismological Society of America},
   pages = {535-553},
   publisher = {Seismological Society of America},
   title = {Amplitudes of horizontally refracted Love waves},
   volume = {64},
   year = {1974},
}

@article{Gregersen1976,
   author = {S Gregersen and L E Alsop},
   issue = {6},
   journal = {Bull. Seismol. Soc. Am.},
   pages = {1855-1872},
   publisher = {Seismological Society of America},
   title = {Mode conversion of Love waves at a continental margin},
   volume = {66},
   year = {1976},
}

@article{Mendiguren1977,
  author  = {Mendiguren, Jorge A.},
  title   = {Inversion of Surface Wave Data in Source Mechanism Studies},
  journal = {Journal of Geophysical Research},
  year    = {1977},
  volume  = {82},
  number  = {5},
  pages   = {889--894},
  doi     = {10.1029/JB082i005p00889}
}

@article{AboZena1979,
  author  = {Abo-Zena, A.},
  title   = {{Dispersion Function Computations for Unlimited Frequency Values}},
  journal = {Geophysical Journal International},
  year    = {1979},
  volume  = {58},
  number  = {1},
  pages   = {91--105}
}

@incollection{Woodhouse1980,
  author    = {Woodhouse, J. H.},
  title     = {{Efficient and Stable Methods for Performing Seismic Calculations in Stratified Media}},
  booktitle = {Proceedings of the International School of Physics ``Enrico Fermi'', Course 78: The Solution of the Inverse Problem in Geophysical Interpretation},
  editor    = {Cassinis, R.},
  publisher = {North-Holland},
  address   = {Amsterdam},
  year      = {1980},
  pages     = {127--151}
}

@article{Dziewonski1981,
   author = {Adam M Dziewonski and Don L Anderson},
   issue = {4},
   journal = {Phys. Earth Planet. In.},
   pages = {297-356},
   publisher = {Elsevier},
   title = {Preliminary reference Earth model},
   volume = {25},
   year = {1981},
}

@article{Kennett1984a,
   author = {B L N Kennett},
   issue = {1},
   journal = {Geophys. J. Int.},
   pages = {235-255},
   publisher = {Oxford University Press},
   title = {Guided wave propagation in laterally varying media–I. Theoretical development},
   volume = {79},
   year = {1984},
}

@article{Its1985,
   author = {E N Its and T B Yanovskaya},
   issue = {1},
   journal = {Wave motion},
   pages = {79-94},
   publisher = {Elsevier},
   title = {Propagation of surface waves in a half-space with vertical, inclined or curved interfaces},
   volume = {7},
   year = {1985},
}

@article{Odom1986,
   author = {Robert I Odom},
   issue = {2},
   journal = {Geophys. J. Int.},
   pages = {425-453},
   publisher = {Oxford University Press},
   title = {A coupled mode examination of irregular waveguides including the continuum spectrum},
   volume = {86},
   year = {1986},
}

@article{Maupin1987,
   author = {V Maupin and B L N Kennett},
   issue = {3},
   journal = {Geophys. J. Int.},
   pages = {837-851},
   publisher = {Oxford University Press},
   title = {On the use of truncated modal expansions in laterally varying media},
   volume = {91},
   year = {1987},
}

@article{Gomberg1988,
   author = {Joan S Gomberg and T Guy Masters},
   doi = {10.1111/j.1365-246X.1988.tb05896.x},
   issn = {1365246X},
   issue = {2},
   journal = {Geophysical Journal},
   pages = {193-218},
   publisher = {Oxford University Press},
   title = {Waveform modelling using locked‐mode synthetic and differential seismograms: application to determination of the structure of Mexico},
   volume = {94},
   year = {1988},
}

@article{Buchen1996,
   author = {P W Buchen and R Ben-Hador},
   issue = {3},
   journal = {Geophys. J. Int.},
   pages = {869-887},
   publisher = {Oxford University Press},
   title = {Free-mode surface-wave computations},
   volume = {124},
   year = {1996},
}

@article{Meier1997a,
   author = {T Meier and P G Malischewsky and H Neunhöfer},
   issue = {6},
   journal = {Bull. Seismol. Soc. Am.},
   pages = {1648-1661},
   publisher = {Seismological Society of America},
   title = {Reflection and transmission of surface waves at a vertical discontinuity and imaging of lateral heterogeneity using reflected fundamental Rayleigh waves},
   volume = {87},
   year = {1997},
}

@article{Meier2000,
   author = {T Meier and P G Malischewsky},
   doi = {10.1046/j.1365-246X.2000.00040.x},
   issn = {0956540X},
   issue = {1},
   journal = {Geophysical Journal International},
   pages = {12-24},
   publisher = {Oxford University Press},
   title = {Approximation of surface wave mode conversion at a passive continental margin by a mode-matching technique},
   volume = {141},
   year = {2000},
}

@book{Aki2002,
   author = {K Aki and P.~G. Richards},
   journal = {Quantitative Seismology, 2nd Ed., by Keiiti Aki and Paul G.~Richards.~Published by University Science Books, ISBN 0-935702-96-2, 704pp, 2002.},
   publisher = {University Science Books},
   title = {Quantitative Seismology, 2nd Ed.},
   year = {2002},
}

@incollection{Snieder2002,
  author    = {Snieder, Roel},
  title     = {Scattering of Surface Waves},
  booktitle = {Scattering and Inverse Scattering in Pure and Applied Science},
  editor    = {Pike, R. and Sabatier, P.},
  pages     = {562--577},
  year      = {2002},
  publisher = {Academic Press},
  address   = {San Diego, California}
}

@article{Chapman2003,
   author = {C H Chapman},
   doi = {10.1046/j.1365-246X.2003.01958.x},
   issn = {0956540X},
   issue = {1},
   journal = {Geophysical Journal International},
   pages = {212-223},
   publisher = {Oxford University Press},
   title = {Yet another elastic plane-wave, layer-matrix algorithm},
   volume = {154},
   year = {2003},
}

@article{Liu2010,
   author = {Tianyun Liu},
   doi = {10.1785/0120090331},
   issn = {00371106},
   issue = {5 A},
   journal = {Bulletin of the Seismological Society of America},
   pages = {2310-2316},
   publisher = {Seismological Society of America},
   title = {{Efficient reformulation of the Thomson-Haskell method for computation of surface waves in layered half-space}},
   volume = {100},
   year = {2010},
}

@article{Denolle2012,
   author = {Marine A Denolle and Eric M Dunham and Gregory C Beroza},
   doi = {10.1785/0120110183},
   issn = {00371106},
   issue = {3},
   journal = {Bulletin of the Seismological Society of America},
   pages = {1214-1223},
   publisher = {Seismological Society of America},
   title = {Solving the surface-wave eigenproblem with {Chebyshev} spectral collocation},
   volume = {102},
   year = {2012},
}

@article{Lin2014,
   author = {Fan Chi Lin and Victor C Tsai and Brandon Schmandt},
   doi = {10.1093/gji/ggu160},
   issn = {1365246X},
   issue = {2},
   journal = {Geophysical Journal International},
   pages = {656-670},
   publisher = {Oxford University Press},
   title = {3-D crustal structure of the western United States: Application of Rayleigh-wave ellipticity extracted from noise cross-correlations},
   volume = {198},
   year = {2014},
}

@article{Datta2017,
   author = {Arjun Datta and Keith F Priestley and Steve Roecker and Chris H Chapman},
   issue = {2},
   journal = {Geophys. J. Int.},
   pages = {1099-1120},
   publisher = {Oxford University Press},
   title = {Surface wave mode coupling and the validity of the path average approximations in surface waveform inversions: an empirical assessment},
   volume = {211},
   year = {2017},
}

@article{Sager2018,
   author = {Korbinian Sager and Laura Ermert and Christian Boehm and Andreas Fichtner},
   doi = {10.1093/gji/ggx429},
   issn = {0956-540X},
   issue = {1},
   journal = {Geophysical Journal International},
   month = {1},
   pages = {566-590},
   publisher = {Oxford University Press},
   title = {Towards full waveform ambient noise inversion},
   volume = {212},
   url = {http://academic.oup.com/gji/article/212/1/566/4411809},
   year = {2018},
}

@article{Fichtner2017,
   author = {Andreas Fichtner and Laurent Stehly and Laura Ermert and Christian Boehm},
   doi = {10.1093/gji/ggw420},
   issn = {0956-540X},
   issue = {2},
   journal = {Geophysical Journal International},
   month = {2},
   pages = {603-638},
   publisher = {Oxford University Press},
   title = {Generalized interferometry – I: theory for interstation correlations},
   volume = {208},
   url = {https://academic.oup.com/gji/article-lookup/doi/10.1093/gji/ggw420},
   year = {2017},
}

@article{Datta2018,
   author = {Arjun Datta},
   doi = {10.5194/gi-7-101-2018},
   issn = {2193-0864},
   issue = {1},
   journal = {Geoscientific Instrumentation, Methods and Data Systems},
   month = {3},
   pages = {101-112},
   title = {{SWRT: A package for semi-analytical solutions of surface wave propagation, including mode conversion, across transversely aligned vertical discontinuities}},
   volume = {7},
   url = {https://www.geosci-instrum-method-data-syst.net/7/101/2018/},
   year = {2018},
}

@article{Ma2016,
   author = {Yiran Ma and Robert W. Clayton and Dunzhu Li},
   doi = {10.1093/gji/ggw235},
   issn = {1365246X},
   issue = {3},
   journal = {Geophysical Journal International},
   pages = {1634-1644},
   title = {Higher-mode ambient-noise Rayleigh waves in sedimentary basins},
   volume = {206},
   year = {2016},
}

@article{Xu2017,
   author = {Zongbo Xu and T. Dylan Mikesell},
   doi = {10.1093/gji/ggx228},
   issn = {0956-540X},
   issue = {3},
   journal = {Geophysical Journal International},
   month = {9},
   pages = {1388-1393},
   publisher = {Oxford University Press},
   title = {On the reliability of direct Rayleigh-wave estimation from multicomponent cross-correlations},
   volume = {210},
   url = {https://academic.oup.com/gji/article-lookup/doi/10.1093/gji/ggx228},
   year = {2017},
}

@article{Hobiger2014,
   author = {Manuel Hobiger and Ulrich Wegler and Katsuhiko Shiomi and Hisashi Nakahara},
   doi = {10.1093/gji/ggu115},
   issn = {1365-246X},
   issue = {1},
   journal = {Geophysical Journal International},
   month = {7},
   pages = {90-109},
   publisher = {Oxford University Press},
   title = {Single-station cross-correlation analysis of ambient seismic noise: application to stations in the surroundings of the 2008 Iwate-Miyagi Nairiku earthquake},
   volume = {198},
   url = {http://academic.oup.com/gji/article/198/1/90/604971/Singlestation-crosscorrelation-analysis-of-ambient},
   year = {2014},
}

@article{Datta2019,
   author = {Arjun Datta},
   doi = {10.1007/s10950-018-9803-4},
   issn = {1383-4649},
   issue = {2},
   journal = {Journal of Seismology},
   month = {3},
   pages = {243-260},
   publisher = {Springer Netherlands},
   title = {On the application of the fk-MUSIC method to measurement of multimode surface wave phase velocity dispersion from receiver arrays},
   volume = {23},
   url = {http://link.springer.com/10.1007/s10950-018-9803-4},
   year = {2019},
}

@article{Tromp2010,
   author = {Jeroen Tromp and Yang Luo and Shravan Hanasoge and Daniel Peter},
   doi = {10.1111/j.1365-246X.2010.04721.x},
   issn = {0956540X},
   issue = {2},
   journal = {Geophysical Journal International},
   month = {11},
   pages = {791-819},
   publisher = {Narnia},
   title = {Noise cross-correlation sensitivity kernels},
   volume = {183},
   url = {https://academic.oup.com/gji/article-lookup/doi/10.1111/j.1365-246X.2010.04721.x},
   year = {2010},
}

@article{Xu2019a,
   author = {Zongbo Xu and T Dylan Mikesell and Gabriel Gribler and Aurélien Mordret},
   doi = {10.1093/gji/ggz261},
   issn = {0956-540X},
   issue = {3},
   journal = {Geophysical Journal International},
   month = {9},
   pages = {1761-1780},
   publisher = {Narnia},
   title = {Rayleigh-wave multicomponent cross-correlation-based source strength distribution inversion. Part 1: Theory and numerical examples},
   volume = {218},
   url = {https://academic.oup.com/gji/article/218/3/1761/5510447},
   year = {2019},
}

@article{Nayak2020,
   author = {Avinash Nayak and Clifford H Thurber},
   doi = {10.1093/gji/ggaa270},
   issn = {0956-540X},
   issue = {3},
   journal = {Geophysical Journal International},
   title = {Using multicomponent ambient seismic noise cross-correlations to identify higher mode Rayleigh waves and improve dispersion measurements},
   volume = {222},
   year = {2020},
}

@article{Xu2020,
   author = {Zongbo Xu and T Dylan Mikesell and Josefine Umlauft and Gabriel Gribler},
   doi = {10.1093/gji/ggaa284},
   issn = {0956-540X},
   issue = {3},
   journal = {Geophysical Journal International},
   month = {9},
   pages = {2084-2101},
   publisher = {Oxford University Press (OUP)},
   title = {Rayleigh-wave multicomponent crosscorrelation-based source strength distribution inversions. Part 2: a workflow for field seismic data},
   volume = {222},
   url = {https://academic.oup.com/gji/article/222/3/2084/5855990},
   year = {2020},
}

@article{Wang2019b,
   author = {Kai Wang and Qinya Liu and Yingjie Yang},
   doi = {10.1029/2018JB017020},
   issn = {21699356},
   issue = {6},
   journal = {Journal of Geophysical Research: Solid Earth},
   month = {6},
   pages = {5794-5810},
   publisher = {Blackwell Publishing Ltd},
   title = {Three-Dimensional Sensitivity Kernels for Multicomponent Empirical Green's Functions From Ambient Noise: Methodology and Application to Adjoint Tomography},
   volume = {124},
   url = {https://onlinelibrary.wiley.com/doi/abs/10.1029/2018JB017020},
   year = {2019},
}

@article{Malkoti2021,
   author = {Ajay Malkoti and Arjun Datta and Shravan M. Hanasoge},
   doi = {10.1093/GJI/GGAB228},
   issn = {0956-540X},
   issue = {1},
   journal = {Geophysical Journal International},
   month = {8},
   pages = {472-482},
   publisher = {Oxford Academic},
   title = {Rayleigh-wave H/V ratio measurement from ambient noise cross-correlations and its sensitivity to VP: a numerical study},
   volume = {227},
   url = {https://academic.oup.com/gji/article/227/1/472/6296642},
   year = {2021},
}

@article{Datta2023b,
   author = {Arjun Datta},
   doi = {10.1093/GJI/GGAD179},
   issn = {0956-540X},
   issue = {3},
   journal = {Geophysical Journal International},
   month = {4},
   pages = {1965-1969},
   publisher = {Oxford Academic},
   title = {{Reply to comment on Malkoti et al. (2021) by Haney and Nakahara}},
   volume = {234},
   url = {https://academic.oup.com/gji/article/234/3/1965/7146201},
   year = {2023},
}

@article{Tsai1971,
  author  = {Tsai, Yih-Ben and Aki, Keiiti},
  title   = {{Amplitude Spectra of Surface Waves from Small Earthquakes and Underground Nuclear Explosions}},
  journal = {Journal of Geophysical Research},
  year    = {1971},
  volume  = {76},
  number  = {17},
  pages   = {3940--3952},
  doi     = {10.1029/JB076i017p03940}
}

@article{Tsai1970,
  author  = {Tsai, Yih-Ben and Aki, Keiiti},
  title   = {{Precise Focal Depth Determination from Amplitude Spectra of Surface Waves}},
  journal = {Journal of Geophysical Research},
  year    = {1970},
  volume  = {75},
  number  = {29},
  pages   = {5729--5744},
  doi     = {10.1029/JB075i029p05729}
}

@article{Romanowicz1984,
  author  = {Romanowicz, Barbara and Guillemant, Philippe},
  title   = {{An Experiment in the Retrieval of Depth and Source Mechanism of Large Earthquakes Using Very Long-Period Rayleigh Wave Data}},
  journal = {Bulletin of the Seismological Society of America},
  year    = {1984},
  volume  = {74},
  number  = {2},
  pages   = {417--437}
}

@article{Datta2026FWANSI2D,
  author  = {Datta, Arjun and Mahesh, Aileni and Lahon, Pankaj},
  title   = {{FW-ANSI2D: A Python package for seismic ambient noise cross-correlation modelling and source inversion}},
  journal = {Journal of Seismology},
  year    = {2026},
  volume  = {30},
  number  = {70},
  doi     = {10.1007/s10950-026-10421-4},
  url     = {https://doi.org/10.1007/s10950-026-10421-4}
}

@article{Herrmann2013,
  author = {Herrmann, Robert B.},
  title = {Computer Programs in Seismology: An Evolving Tool for Instruction and Research},
  journal = {Seismological Research Letters},
  volume = {84},
  number = {6},
  pages = {1081--1088},
  year = {2013},
  doi = {10.1785/0220110096},
}

@article{Ekstrom2012,
  author  = {Ekstr{\"o}m, G{\"o}ran and Nettles, Meredith and Dziewo{\'n}ski, Adam M.},
  title   = {The global {CMT} project 2004--2010: Centroid-moment tensors for 13,017 earthquakes},
  journal = {Physics of the Earth and Planetary Interiors},
  volume  = {200--201},
  pages   = {1--9},
  year    = {2012},
  doi     = {10.1016/j.pepi.2012.04.002}
}

@article{Molnar1969,
  author  = {Molnar, Peter and Savino, John and Sykes, Lynn R. and
             Liebermann, Robert C. and Hade, George and Pomeroy, Paul W.},
  title   = {Small Earthquakes and Explosions in Western North America
             Recorded by New High Gain, Long Period Seismographs},
  journal = {Nature},
  volume  = {224},
  number  = {5226},
  pages   = {1268--1273},
  year    = {1969},
  doi     = {10.1038/2241268a0}
}

@article{Saito1967,
  author  = {Saito, Masanori},
  title   = {Excitation of free oscillations and surface waves by a point source in a vertically heterogeneous Earth},
  journal = {Journal of Geophysical Research},
  year    = {1967},
  volume  = {72},
  number  = {14},
  pages   = {3689--3699},
  doi     = {10.1029/JZ072i014p03689}
}

@article{Brissaud2020,
  author  = {Brissaud, Quentin and Bowden, Daniel C. and Tsai, Victor C.},
  title   = {Extension of the Basin Rayleigh-Wave Amplification Theory to Include Basin-Edge Effects},
  journal = {Bulletin of the Seismological Society of America},
  year    = {2020},
  volume  = {110},
  number  = {3},
  pages   = {1305--1322},
  doi     = {10.1785/0120190161}
}

@phdthesis{Datta2016,
  author  = {Datta, Arjun},
  title   = {Surface wave mode coupling due to lateral heterogeneity and its impact on waveform inversions},
  school  = {University of Cambridge},
  year    = {2016},
  month   = Dec,
  address = {Cambridge, UK}
}

@techreport{Woodhouse1981,
  author      = {Woodhouse, J. H.},
  title       = {The Calculation of the Complete Normal Mode Spectrum of the Earth Using Finite Difference Methods},
  institution = {Applied Seismology Group, Lincoln Laboratory, MIT},
  address     = {Lexington, Massachusetts},
  type        = {Semi-annual Technical Summary},
  year        = {1981}
}

\end{document}